\documentclass{PoS}
\usepackage{rotating}
\usepackage{multirow}
\usepackage{tabulary}
\usepackage{dcolumn}
\usepackage{lipsum}
\usepackage{bibentry}
\usepackage{mhchem} 
\usepackage{booktabs}

\title{Phenomenology of Neutrinoless Double Beta Decay}

\ShortTitle{Phenomenology of Neutrinoless Double Beta Decay}

\author{\speaker{J.J.\ G\'omez-Cadenas} and Justo Mart\'in-Albo\\
Instituto de F\'isica Corpuscular (IFIC), CSIC \& Universitat de Valencia\\
Calle Catedr\'atico Jos\'e Beltr\'an, 2, 46980 Paterna, Valencia, Spain\\
E-mail: \email{gomez@mail.cern.ch}, \email{jmalbos@ific.uv.es}}

\abstract{This paper reviews the current status and future outlook of neutrinoless double beta decay searches, which try to provide an answer to the fundamental question of whether neutrinos are Dirac or Majorana particles.}

\FullConference{Gran Sasso Summer Institute 2014 Hands-On Experimental Underground Physics at LNGS - GSSI14,\\
		22 September - 03 October 2014\\
		INFN - Laboratori Nazionali del Gran Sasso, Assergi, Italy }

\newcommand{\bb}{\ensuremath{\beta\beta}}
\newcommand{\bbonu}{\ensuremath{0\nu\beta\beta}}
\newcommand{\bbtnu}{\ensuremath{2\nu\beta\beta}}

\newcommand{\mbb}{\ensuremath{m_{\bb}}}
\newcommand{\Tonu}{\ensuremath{T_{1/2}^{0\nu}}}
\newcommand{\Gonu}{\ensuremath{G^{0\nu}}}
\newcommand{\Monu}{\ensuremath{\left|M^{0\nu}\right|}}
\newcommand{\Qbb}{\ensuremath{Q_{\bb}}}

\newcommand{\ckky}{\ensuremath{\mathrm{cts~keV^{-1}~kg^{-1}~yr^{-1}}}}
\newcommand{\CA}{\ensuremath{^{48}\mathrm{Ca}}}
\newcommand{\GE}{\ensuremath{^{76}\mathrm{Ge}}}
\newcommand{\SE}{\ensuremath{^{82}\mathrm{Se}}}
\newcommand{\MO}{\ensuremath{^{100}\mathrm{Mo}}}
\newcommand{\CD}{\ensuremath{^{116}\mathrm{Cd}}}
\newcommand{\TE}{\ensuremath{^{130}\mathrm{Te}}}
\newcommand{\XE}{\ensuremath{^{136}\mathrm{Xe}}}
\newcommand{\ND}{\ensuremath{^{150}\mathrm{Nd}}}

\newcommand{\TL}{\ensuremath{^{208}\mathrm{Tl}}}
\newcommand{\BI}{\ensuremath{^{214}\mathrm{Bi}}}

\newcommand{\NA}{\ensuremath{^{22}}Na}
\newcommand{\CS}{\ensuremath{^{137}}Cs}

\newcommand{\hair}{\ifmmode\mskip1mu\else\kern0.08em\fi}

\begin{document}

%%%%%%%%%%%%%%%%%%%%%%%%%%%%%%%%%%%%%%%%%%%%%%%%%%%%%%%%%%%%
\section{Introduction}
%%%
Neutrinoless double beta (\bbonu) decay is a postulated very slow radioactive process in which two neutrons inside a nucleus transform into two protons emitting two electrons. The discovery of this process would demonstrate that neutrinos are Majorana particles and that total lepton number is not conserved in nature, two findings with far-reaching implications in particle physics and cosmology. First, the existence of Majorana neutrinos implies a new energy scale at a level inversely proportional to the observed neutrino masses. Such a scale, besides providing a simple explanation for the striking lightness of neutrino masses, is probably connected to several open questions in particle physics, like the origin of mass or the flavour problem. Second, Majorana neutrinos violate the conservation of lepton number, and this, together with CP violation, could be responsible, through the mechanism known as leptogenesis, for the observed cosmological asymmetry between matter and antimatter. 

After 75 years of experimental effort, no compelling evidence for the existence of \bbonu\ decay has been obtained, but a new generation of experiments that are already running or about to run promises to push forward the current limits exploring the degenerate-hierarchy region of neutrino masses. In order to do that, the experiments are using masses of \bbonu\ isotopes ranging from tens of kilograms to several hundreds, and will need to improve the background rates achieved by previous experiments by, at least, an order of magnitude. If no signal is found, masses of the order of thousands of kilograms and further background reduction will be required. In spite of the formidable experimental challenge (or possibly because of it), the field is brimming with new ideas. However, the requirements are often conflicting, and none of the considered technologies is capable of simultaneously optimizing all parameters. 

This paper summarizes the current status of the field and makes an attempt to define the best strategy for the exploration of the inverted-hierarchy region of neutrino masses.\footnote{Many other reviews have been published on this topic over the last few years; see, for instance, Refs.~\cite{GomezCadenas:2011it, Elliott:2012sp, Giuliani:2012zu, Cremonesi:2013vla, Bilenky:2014uka}.} The diversity of experimental approaches we are currently witnessing will not be viable at that scale, and only two or three technologies (most likely based on different isotopes) are going to be retained. Their suitability for the task depends essentially on their scalability to large masses ---\thinspace including here the possible difficulties in the procurement and isotopic enrichment of tonnes of material\thinspace--- and on their capability to control the backgrounds to the required extremely low levels, which will have to be demonstrated in the current-generation experiments.

%%%%%%%%%%%%%%%%%%%%%%%%%%%%%%%%%%%%%%%%%%%%%%%%%%%%%%%%%%%%
\section{Neutrinoless double beta decay and Majorana neutrinos}
%%%
Double beta (\bb) decay is a second-order weak process that transforms a nuclide of atomic number $Z$ into its isobar with atomic number $Z+2$. The ordinary decay mode consisting in two simultaneous beta decays (\bbtnu),
%%%
\begin{equation}
(Z,A) \to (Z+2,A) + 2~e^{-} + 2~\overline{\nu}_{e}\, ,
\end{equation}
%%%
was first considered by Maria Goeppert-Mayer, in 1935 \cite{GoeppertMayer:1935qp}. There has been geochemical evidence of its existence since the 1950s \cite{Inghram:1950qv}, but the first direct observation, in \SE\ and using a time projection chamber as detector, was not made until 1987 \cite{Elliott:1987kp}. Since then, it has been repeatedly observed in several nuclides with typical lifetimes of the order of $10^{18}$--$10^{21}$ years, the longest ever measured among radioactive decay processes. With such long half-lives, for \bbtnu\ to be a competitive decay mode, the $\beta$ decay to the $Z+1$ nuclide must be either energetically forbidden or highly suppressed. Such a condition is fulfilled by 35 naturally-occurring isotopes thanks to the nuclear pairing force, which ensures that even-even nuclides are more bound than their odd-odd isobars.

The neutrinoless decay mode (\bbonu),
\begin{equation}
(Z,A) \rightarrow (Z+2,A) + 2\ e^{-}, \label{eq:bb0nu}
\end{equation}
was proposed by Wendell H.\ Furry in 1939 \cite{Furry:1939qr} as a method to test Majorana's theory~\cite{Majorana:1937vz} applied to neutrinos. In contrast to the two-neutrino mode, the neutrinoless mode violates total lepton number conservation, and is, therefore, forbidden in the Standard Model of particle physics. Its existence is linked to that of Majorana neutrinos \cite{Schechter:1980gr}. No convincing experimental evidence of the decay exists to date. 

%%%%%%%%%%
\begin{figure}
\centering
\includegraphics[width=0.55\textwidth]{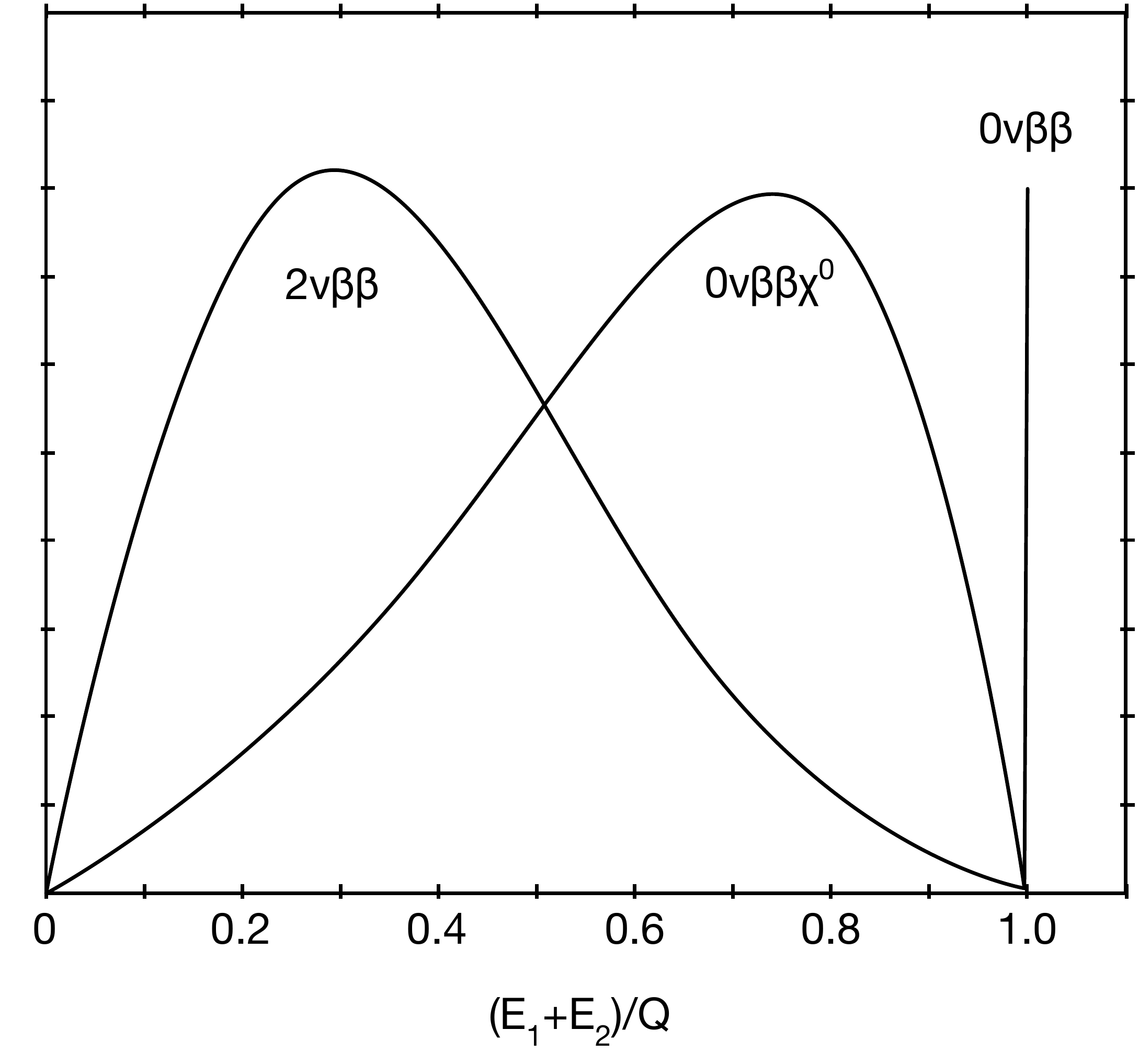}
\caption{Spectra for the sum of the kinetic energies of the two emitted electrons  in three different \bb\ modes: \bbtnu, \bbonu\ and \bb\ decay with Majoron emission. The amplitudes are arbitrary.} \label{fig:BBModes}
\end{figure}
%%%%%%%%%%

Phase-space considerations alone would give preference to the \bbonu\ mode over the \bbtnu\ one, but the decay rate of the former is suppressed by the very small neutrino masses. Both transition modes involve the $0^{+}$ ground state of the initial nucleus and, in almost all cases, the $0^{+}$ ground state of the final nucleus. For some isotopes, it is also energetically possible to have a transition to an excited $0^{+}$ or $2^{+}$ final state, even though these are suppressed because of the smaller phase space available. In both decay modes the emitted leptons carry essentially all the available energy and the nuclear recoil is negligible. Therefore, in the \bbonu\ mode, the spectrum for the sum of the kinetic energies of the emitted electrons (see Figure~\ref{fig:BBModes}) is a mono-energetic line at $\Qbb$, the $Q$ value of the reaction, defined as the mass difference between the parent and daughter nuclides:
%%%
\begin{equation}
Q_{\bb} \equiv M(A, Z) - M(A, Z+2).
\end{equation}
%%%
In the case of the \bbtnu-decay mode, the spectrum is continuous, extending from 0 to \Qbb\ and peaking below $\Qbb/2$. In addition to the the two basic decay modes described above, several decay modes involving the emission of a light neutral boson, the Majoron ($\chi^{0}$), have been proposed in extensions of the Standard Model. 

%%%%%%%%%%
\begin{figure}
\centering
\includegraphics[scale=0.625]{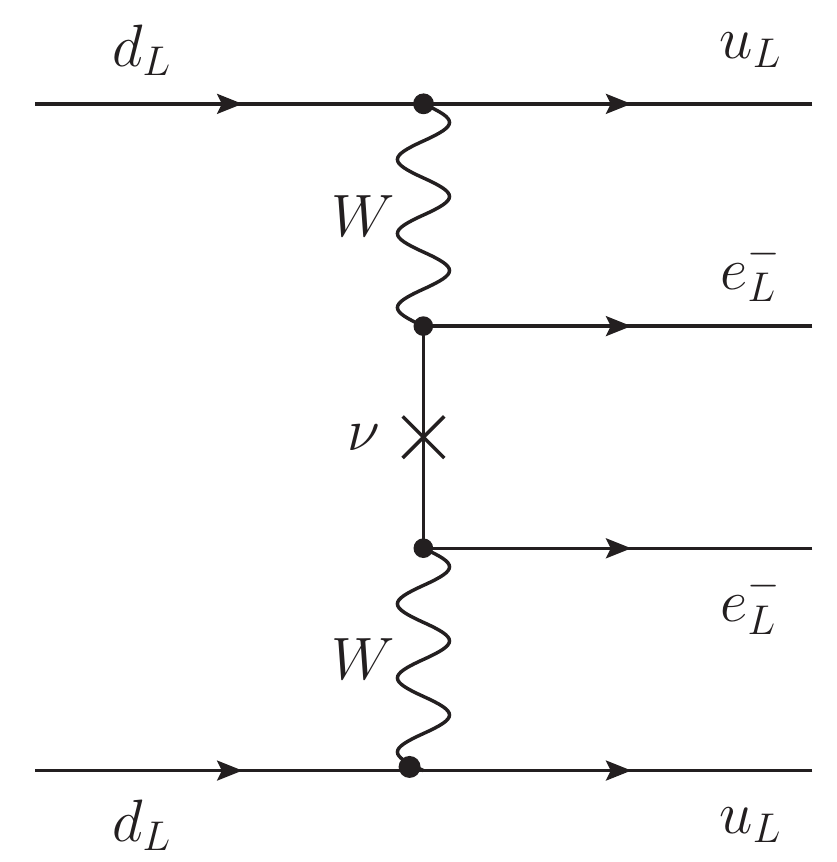}
\caption{Neutrinoless double beta decay mediated by the standard mechanism, the virtual exchange of a light Majorana neutrino.} \label{fig:LightMajoranaNuExchange}
\end{figure}
%%%%%%%%%%

The simplest underlying mechanism inducing neutrinoless double beta decay is the exchange of a light Majorana neutrino (see Figure~\ref{fig:LightMajoranaNuExchange}): the parent nucleus emits a pair of virtual $W$ bosons, and then these exchange a Majorana neutrino to produce the outgoing electrons. At the vertex where it is emitted, the exchanged neutrino is created, in association with an electron, as an antineutrino with almost total positive helicity, and only its small, $\mathcal{O}(m_{\nu}/E)$, negative-helicity component is absorbed at the other vertex. Considering that the amplitude is, in this case, a sum over the contributions of the three light neutrino mass states $\nu_i$ and is proportional to $U_{ei}^2$, we conclude that the modulus of the amplitude for the \bbonu\ process must be proportional in this case to the \emph{effective neutrino Majorana mass}:
%%%
\begin{equation}
\mbb \equiv \left|\, \sum_{i=1}^3 U_{ei}^2\cdot m_i\ \right|, \label{eq:mbb}
\end{equation}
%%%
where $U_{ei}$ are the elements of the first row of the neutrino mixing matrix and $m_{i}$ are the three neutrino masses.

Other sources of new physics  beyond the Standard Model (right-handed weak currents, for instance) could also cause neutrinoless double-beta decay \cite{Feinberg:1959, Mohapatra:1980yp, Tello:2010am} and, perhaps, dominate its decay rate. Nevertheless, such models induce Majorana mass for neutrinos from radiative corrections as well.

In the case where light Majorana neutrino exchange is the dominant contribution to \bbonu-decay, the inverse of the half-life for the process can be written as
%%%
\begin{equation}
\left(T^{0\nu}_{1/2}\right)^{-1} = G^{0\nu}\ \left|M^{0\nu}\right|^2\ \left(\frac{\mbb}{m_{e}}\right)^2.
\label{eq:Tonu}
\end{equation}
%%%
Here, \Gonu\ is a phase-space factor that depends on the transition $Q$ value and on the nuclear charge $Z$, and $M^{0\nu}$ is the nuclear matrix element (NME) for the process. The phase-space factor can be calculated analytically with sufficient accuracy (error estimates of about 1 per mille) \cite{Kotila:2012zza}. The NME is evaluated using nuclear models, although with considerable uncertainty (see Section~\ref{sec:NME}). In other words, the value of the effective neutrino Majorana mass, \mbb, can be inferred from a non-zero \bbonu-rate measurement, even though with some nuclear physics uncertainties. Conversely, if a given experiment does not observe the \bbonu\ process, the result can be interpreted in terms of an upper bound on \mbb.  

\begin{figure}
\centering
\includegraphics[width=0.525\textwidth]{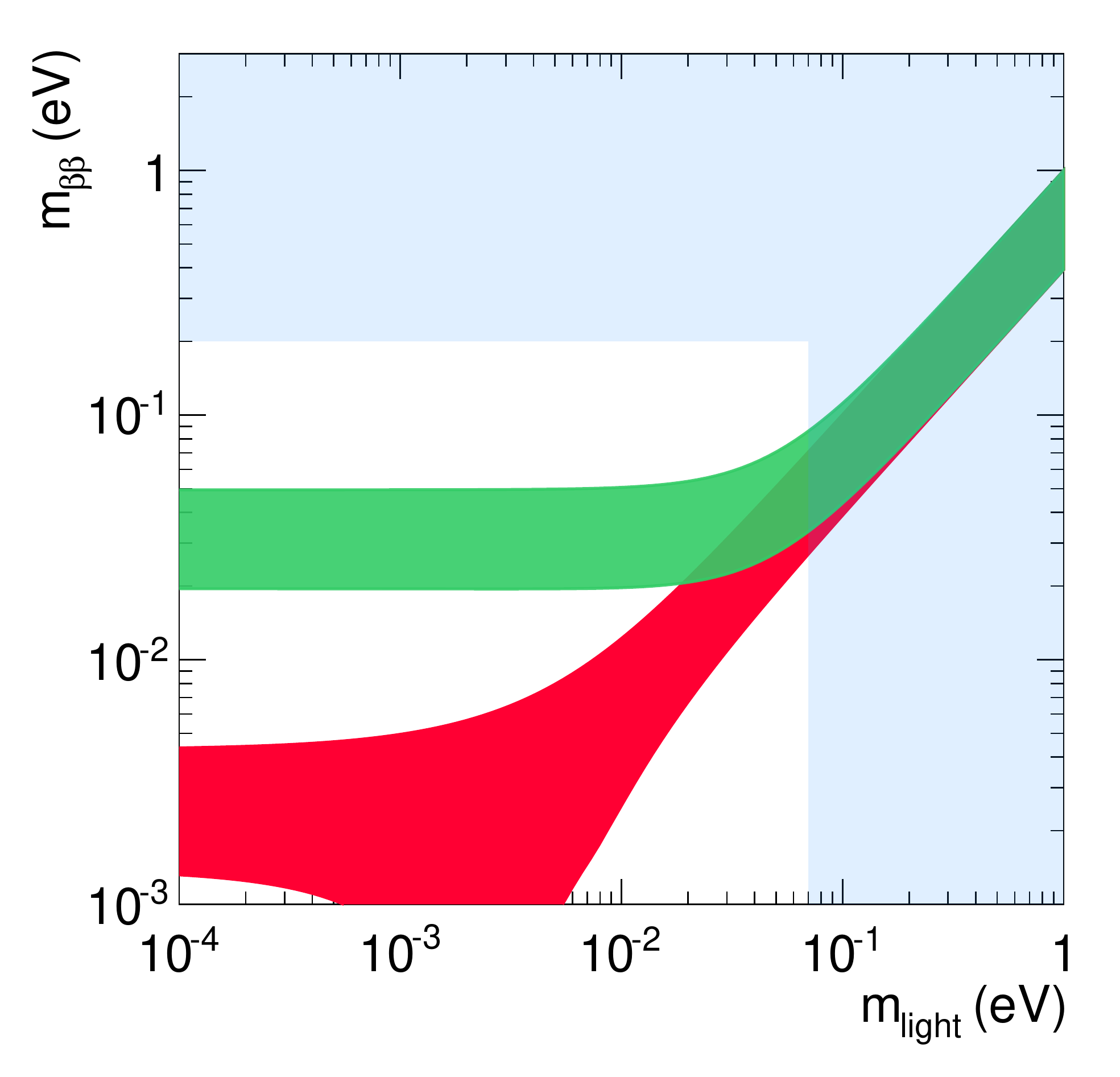}
\caption{The effective neutrino Majorana mass, \mbb, as a function of the lightest neutrino mass, $m_\mathrm{light}$. The green band corresponds to the inverted ordering of neutrino masses ($m_\mathrm{light}\equiv m_{3}$), while the red band corresponds to the normal ordering ($m_\mathrm{light}\equiv m_{1}$). The vertically-excluded region comes from cosmological bounds \cite{Ade:2013zuv}, the horizontal one from \bbonu\ constraints \cite{Albert:2014awa, Asakura:2014lma}. This graphical representation was first proposed by F.~Vissani \cite{Vissani:1999tu}.} \label{fig:BetaBetaVsLight}
\end{figure}

If light Majorana neutrino exchange is the dominant mechanism for \bbonu\ decay, it is clear from Eq.~(\ref{eq:mbb}) that the decay is then directly connected to neutrino oscillations phenomenology, and that it also provides direct information about the absolute neutrino mass scale, as cosmology \cite{Lesgourgues:2012uu} and $\beta$-decay experiments \cite{Drexlin:2013lha} do. The relationship between \mbb\ and the actual neutrino masses $m_i$ is affected by the uncertainties in the measured oscillation parameters, the unknown neutrino mass ordering (normal or inverted) and the unknown phases in the neutrino mixing matrix (both Dirac and Majorana). For example, the relationship between \mbb\ and the lightest neutrino mass, $m_\mathrm{light}$, is shown in Figure~\ref{fig:BetaBetaVsLight}. The width of the two bands is due to the unknown CP violation phases and the uncertainties in the measured oscillation parameters \cite{Gonzalez-Garcia:2014bfa}. Figure \ref{fig:BetaBetaVsLight} also shows the upper bound on $m_\mathrm{light}$ from cosmology ($m_\mathrm{light}<0.07$ eV \cite{Ade:2013zuv}), and an upper bound on \mbb\ from \bbonu-decay searches ($m_{\beta\beta}<0.2$ eV \cite{Albert:2014awa, Asakura:2014lma}). As can be seen from the figure, the bound from \bbonu-decay data on the absolute mass scale is almost as stringent as that from cosmological observations.

%%%%%%%%%%
%%%%%%%%%%

%%%%%%%%%%%%%%%%%%%%%%%%%%%%%%%%%%%%%%%%%%%%%%%%%%%%%%%%%%%%
\section{Nuclear matrix elements} \label{sec:NME}
%%%
All nuclear structure effects of neutrinoless double beta decay are included in the nuclear matrix element (NME). Its knowledge is essential in order to relate a possible half-life measurement to the neutrino masses, and to compare the sensitivity and results of experiments using different \bb\ isotopes. NMEs cannot be separately measured and must be evaluated theoretically. Unfortunately, due to the many-body nature of the nuclear problem, only approximate estimates can be obtained at present time. A variety of techniques are used for this; namely: the \emph{interacting shell model} (ISM) \cite{Caurier:2007wq, Menendez:2008jp}, the \emph{quasiparticle random-phase approximation} (QRPA) \cite{Rodin:2006yk, Kortelainen:2007rn}, the \emph{interacting boson model} (IBM-2) \cite{Barea:2013bz} and the \emph{energy density functional method} (EDF) \cite{Rodriguez:2010mn, Vaquero:2014dna}.\footnote{It is beyond the scope of this work to provide a detailed discussion of these methods. The interested reader may consult, for instance, the recent review paper by P.~Vogel~\cite{Vogel:2012ja} and the specialized references given above.}

%%%%%%%%%%
\begin{figure}
\centering
\includegraphics[width=0.75\textwidth]{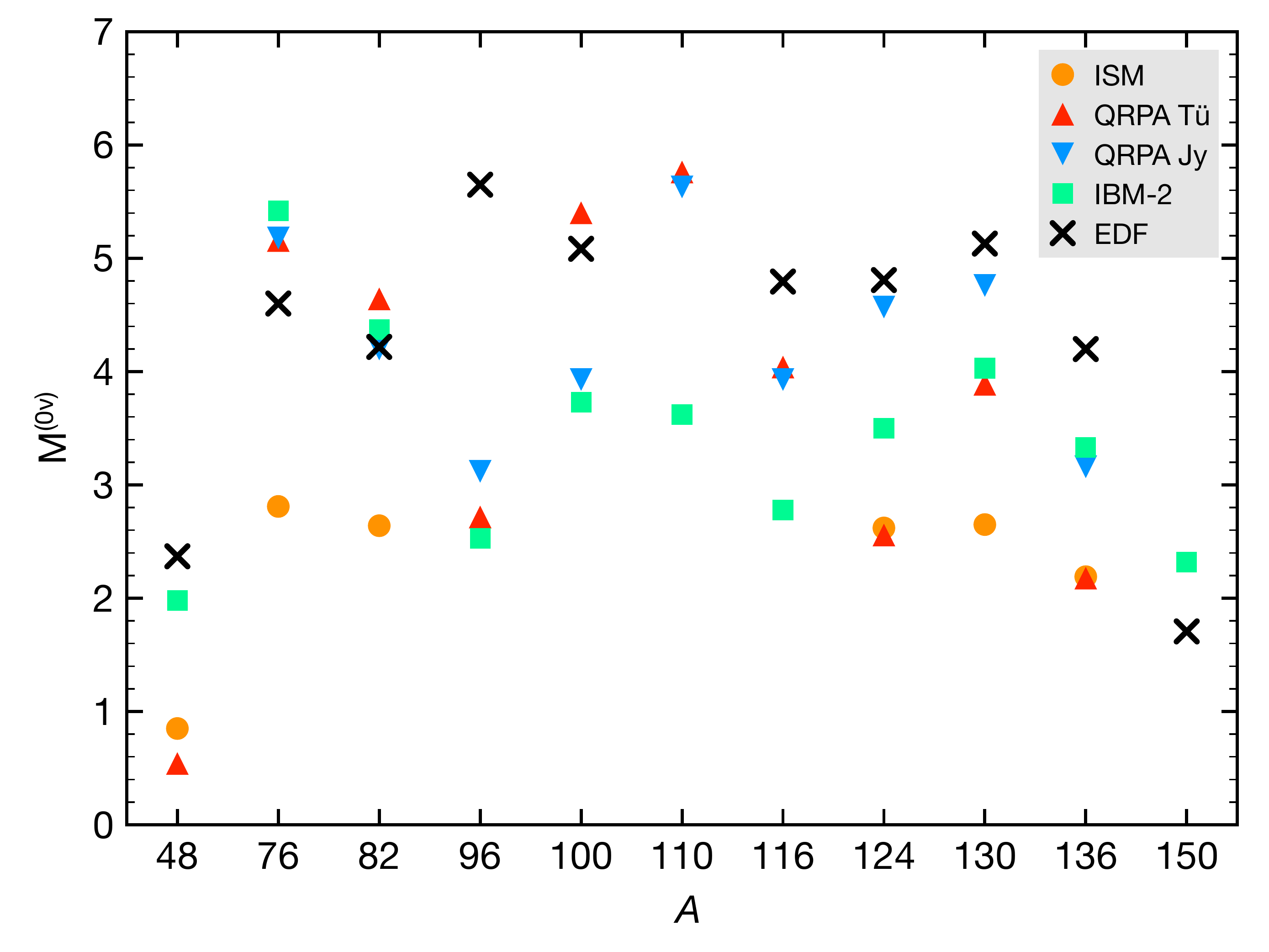}
\caption{Nuclear matrix elements (NMEs) for \bbonu\ decay to the ground state as calculated in four different frameworks: \emph{interacting shell model} (ISM) \cite{Menendez:2008jp}, \emph{quasiparticle random-phase approximation} by the T\"ubingen (QRPA T\"u) \cite{Simkovic:2013qiy} and the Jyv\"askyl\"a (QRPA Jy) \cite{Suhonen:2012ii} groups, \emph{interacting boson model} (IBM-2) \cite{Barea:2013bz} and \emph{energy density functional} method (EDF) \cite{Vaquero:2014dna}. 
The IBM-2 calculation uses the Jastrow Miller-Spencer short-range correlations; ISM, EDF and QRPA Jy use UCOM; and QRPA T\"u uses the Argonne potential. 
IBM-2 and QRPA T\"u NMEs are evaluated with $g_{A}=1.269$, and the rest are evaluated using $g_{A}=1.25$.}\label{fig:NME}
\end{figure}
%%%%%%%%%%

Figure \ref{fig:NME} summarizes the results of the most recent calculations. The reliability of the calculations has greatly improved in the last few years, and although the results from the different techniques are not yet completely convergent, differing by up to a factor of 2, they seem to be at least fairly insensitive to the broad range of approximations made \cite{Vogel:2012ja}. Therefore, if \bbonu\ decay were observed in one nucleus, one would be able to predict its lifetime in a different candidate nuclei with some confidence, increasing the chances that a reliable and confirmed result is obtained. In any case, it is clear that further progress in the calculation of NMEs is needed to reduce the overall theoretical uncertainty. In particular, the origin of the differences between the models should be better understood. Work in that direction has already started \cite{Menendez:2014ena}.

The dependence of the NME on the effective axial-vector coupling constant, $g_{A}$, introduces another significant uncertainty in the calculated rates. Barea et al.\ \cite{Barea:2013bz} and Ejiri \cite{Ejiri:2010zza} have fitted the known half-lives for \bbtnu\ decay and find effective values of $g_{A}$ of about 0.8 for ISM calculations and 0.6 for IBM-2, with a mild dependence on the mass number of the isotope. In contrast, the calculated phase-space factors for neutrinoless decay are generally presented with the free-nucleon value $g_{A} \simeq1.27$ \cite{Agashe:2014kda}, measured in the weak interactions and decays of nucleons. The difference between this value and 0.6 corresponds to a factor of 20 in decay rate. The extent of the renormalization of $g_{A}$ in \bbonu\ decay remains a topic of discussion among nuclear theorists.
 
%%%%%%%%%%%%%%%%%%%%%%%%%%%%%%%%%%%%%%%%%%%%%%%%%%%%%%%%%%%%
\section{The search for neutrinoless \mbox{double beta decay}} 
%%%
The discovery of neutrinoless double beta decay would represent a major breakthrough in particle physics. A single unequivocal observation of the decay would prove the Majorana nature of neutrinos and the violation of total lepton number. Unfortunately, this is by no means an easy task. The design of a detector capable of identifying efficiently and unambiguously such a rare signal poses a considerable experimental problem. To begin with, one needs a large mass of the scarce \bb\ isotopes to probe in a reasonable time the extremely long lifetimes predicted for the process. For instance, for a Majorana neutrino mass of 50~meV, we can estimate using Eq.~(\ref{eq:Tonu}) that half-lives in the range of $10^{26}$ to $10^{27}$ years must be explored, i.e.\ 17 orders of magnitude longer than the age of the universe! 

A better sense of what such long half-lives mean can be grasped with a simple calculation. Consider the radioactive decay law in the approximation $T_{1/2}\gg t$, where $t$ is the observation time. In that case, the expected number of \bbonu\ decays is given by
\begin{equation}
N = \log2\ \frac{\varepsilon\cdot M \cdot N_\mathrm{A}}{W}\ \frac{t}{\Tonu}\ , \label{eq:Nbb}
\end{equation}
where $M$ is the mass of the \bb-emitting isotope, $W$ is its molar mass, $N_\mathrm{A}$ is the Avogadro constant and $\varepsilon$ is the detection efficiency. It follows from the above equation that in order to observe one decay per year, assuming perfect detection efficiency and no disturbing background, and for a Majorana neutrino mass of 50~meV, macroscopic masses of \bb\ isotope of the order of 100~kg are needed.

The situation becomes even more challenging when considering real experimental conditions. The detectors used in double beta decay experiments are designed, in general, to measure the energy of the radiation emitted by a \bb\ source. In a neutrinoless double beta decay the sum of the kinetic energies of the two released electrons is always equal to the $Q$ value of the process. However, due to the finite energy resolution of any detector, \bbonu\ events spread over an energy range centred around \Qbb, typically following a Gaussian distribution. Other processes occurring in the detector can fall within that energy window, becoming a background and compromising drastically the sensitivity of the experiment.

The background processes that can mimic a $\bbonu$-decay signal in a detector are abundant. To begin with, the experiments have to deal with the intrinsic background from the standard two-neutrino double beta decay, which can only be distinguished from the signal by measuring the energy of the emitted electrons, since the neutrinos go undetected. Good energy resolution is, therefore, essential to prevent the \bbtnu\ spectrum tail from spreading over the \bbonu\ peak. However, this \emph{energy signature} is not enough per se, since the continuous energy spectrum arising from natural radioactivity can easily overwhelm the signal peak. For this reason, additional experimental signatures and careful selection of radiopure materials are crucial. 

Several other factors, like the detection efficiency or the scalability to large masses, must be taken into account as well during the design of a double beta decay experiment. The simultaneous optimization of all these parameters is often conflicting, if not impossible, and hence many different experimental techniques have been proposed. In order to compare them, a figure of merit, the experimental \emph{sensitivity} to \mbb, is normally used. 

%%%
The \emph{sensitivity} of an experiment searching for new phenomena is a measure of the result that would be obtained in the absence of a true signal. More precisely, it is defined as the average confidence limit one would get from a large ensemble of experiments with the same expected background and no signal. Accordingly, the sensitivity of a double beta decay experiment to \mbb\ can be expressed combining Equations~(\ref{eq:Tonu}) and (\ref{eq:Nbb}) as
%%%
\begin{equation}
S(\mbb) = A\ \sqrt{\frac{\overline{N}}{\varepsilon Mt}}\,, \label{eq:Sensitivity1}
\end{equation}
%%%
where
%%%
\begin{equation}
A \equiv \left(\frac{W}{N_\mathrm{A}\log2}\ \frac{m_{e}^{2}}{\Gonu\Monu^{2}} \right)^{1/2}
\end{equation}
%%%
is a constant that depends only on the considered \bb\ isotope and $\overline{N}$ is the average upper limit on the expected number of events in the absence of signal.

For an ideal, background-free \bbonu-decay experiment, under the no-signal hypothesis, the observed number of events would always be equal to zero, with no fluctuations. The average upper limit, $\overline{N}$, is in this case simply $(Mt)^{-1/2}$. For an experiment with expected background $b$, the average upper limit in the limit of large background, is well approximated  by the expression:
%%%
\begin{equation}
\overline{N}(b) \simeq k~\sqrt{b}\ . \label{eq:AverageNLimit}
\end{equation}
%%%
Substituting Eq.~(\ref{eq:AverageNLimit}) into Eq.~(\ref{eq:Sensitivity1}), we obtain
%%%
\begin{equation}
S(\mbb) = A'\ \left(\frac{b^{1/2}}{\varepsilon~M~t}\right)^{1/2}\,, \label{eq:Sensitivity2}
\end{equation}
%%%
with
%%%
\begin{equation}
{A}' \equiv {A}\ \sqrt{k}\,.
\end{equation}
%%%
Usually, the background is approximately proportional to the exposure, $Mt$, and to the width of the energy window $\Delta E$ defined by the resolution of the detector:
%%%
\begin{equation}
b = c \cdot M \cdot t \cdot \Delta E \,,
\end{equation}
with the background rate $c$ typically expressed in $\mathrm{counts}/(\mathrm{keV}\cdot\mathrm{kg}\cdot\mathrm{year})$.\footnote{Traditionally, the exposure and background rate have been expressed in units of \emph{detector mass}. However, given the disparity of detector masses used by the new generation of \bbonu-decay experiments (for example, KamLAND-Zen deploys 13 tonnes of \XE-loaded liquid scintillator, whereas GERDA will use about 40~kg of enriched germanium diodes in its seconds phase), in this work, in order to facilitate the comparison between experiments, we will express the relevant quantities in units of \bb\ isotope mass unless stated otherwise.} Equation~(\ref{eq:Sensitivity2}) then becomes
%%%
\begin{equation}
{S}(\mbb) = {A}'\ \sqrt{1/\varepsilon}\ \left(\frac{c~\Delta E}{M~t}\right)^{1/4}\,. \label{eq:Sensitivity3}
\end{equation}
%%%
In short, the background limits dramatically the sensitivity of a double beta decay experiment, improving only as $(Mt)^{-1/4}$ instead of the $(Mt)^{-1/2}$ expected in the background-free case.

%%%%%%%%%%%%%%%%%%%%%%%%%%%%%%%%%%%%%%%%%%%%%%%%%%%%%%%%%%%%
\section{Ingredients for double beta decay experiments} 
%%%
The first objective of the new generation of double beta decay experiments, to confirm or refute experimentally the claim of a \bbonu-decay signal in \GE, has already been tackled by the leading experiments of the current generation,  EXO-200, KamLAND-Zen and GERDA. In the forthcoming years  more data from other experiments will be available as well (see next section). The ultimate goal of all these projects is the exploration of the inverted hierarchy of neutrino masses, a very ambitious objective that will require exposures close to 10$^{4}$~kg~yr and background levels of the order of $10^{-3}$~\ckky\ or better. Given the scale, cost and risk that characterize these experiments, it seems prudent to build as a first step an $\mathcal{O}(100)$-kg detector that demonstrates the performance of the experimental technique. Nevertheless, scaling up these detector to large \bb\ masses will not be straightforward in most cases. 

As we have seen, the sensitivity of any \bbonu-decay experiment depends, ultimately, on a few parameters; namely: energy resolution, background rate, detection efficiency, exposure and \bb\ isotope. In this section, we discuss the requirements that the exploration of the inverted-hierarchy region of neutrino masses imposes on them. 
Some of the parameters (the energy resolution, for instance) depend only on the experimental technique and cannot be improved at will. Others are also determined by factors unrelated to the detection technique (e.g.\ the background rate may depend on the availability of radiopure materials or the depth of the underground laboratory), leaving more room for improvement.

%%%%%%%%%%%%%%%%%%%%%%%%%%%%%%%%%%%%%%%%
\subsection{Choice of the isotope} \label{subsec:Isotope}
\begin{figure}
\centering
\includegraphics[width=0.75\textwidth]{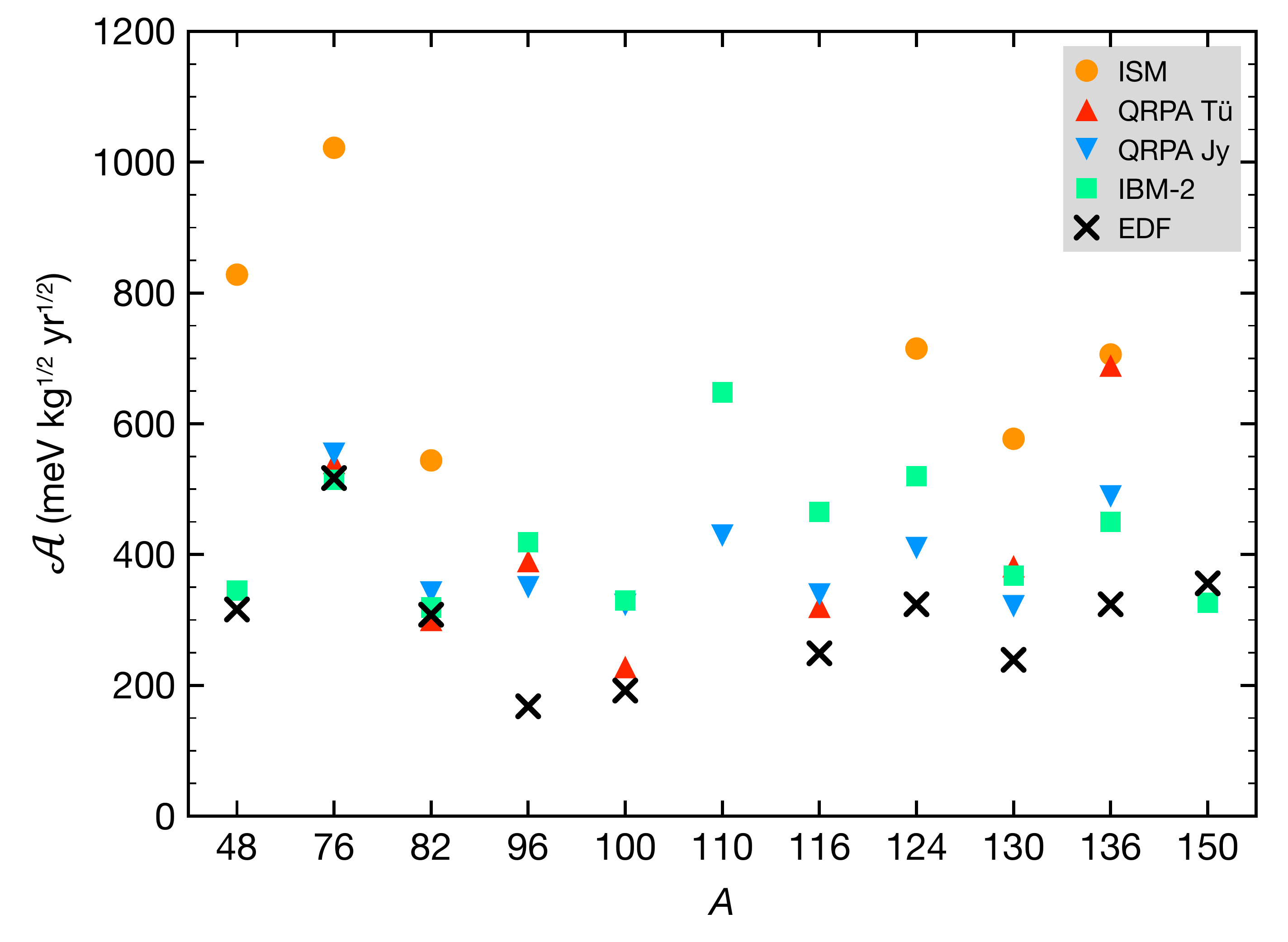}
\caption{Isotopic constant ${A}$  of the eleven \bb\ isotopes
with a $Q$ value larger than 2~MeV. Variations of about a factor of 2 can be found for a given set of NMEs.} \label{fig:IsotopicConstant}
\end{figure}

%%%
Thirty-five naturally-occurring isotopes are \bb\ emitters. Which ones are the most adequate for neutrinoless double beta decay searches? Let us start with considerations of the most favourable phase-space factors and nuclear matrix elements. To a first approximation, the phase-space factor \Gonu\ varies as $\Qbb^5$ \cite{Kotila:2012zza}. Isotopes with large $Q$ values are therefore strongly favoured. For this reason, only the eleven with $\Qbb>2$~MeV have usually been considered for \bbonu-decay searches. As for the NMEs, all calculation methods agree that the value of \Monu\ does not change abruptly from one candidate isotope to another one. The isotopic constant ${A}$ defined in Eq.~(\ref{eq:Sensitivity1}) shows variations of about a factor of 2 (see Figure~\ref{fig:IsotopicConstant}). In other words, within a factor of 2, the decay rate per unit mass does not depend on the \bb\ isotope.

%%%%%%%%%%
%%%%%%%%%%

Another advantage in choosing a \bb\ isotope with a high $Q$ value comes in the form of background control. As we will discuss later, backgrounds from natural radioactivity populate the energy region below 3~MeV. The possibility to use an isotope with $Q_{\beta\beta}$ above these energies is therefore desirable.

%%%%%%%%%%
\begin{figure}
\centering
\includegraphics[width=0.55\textwidth]{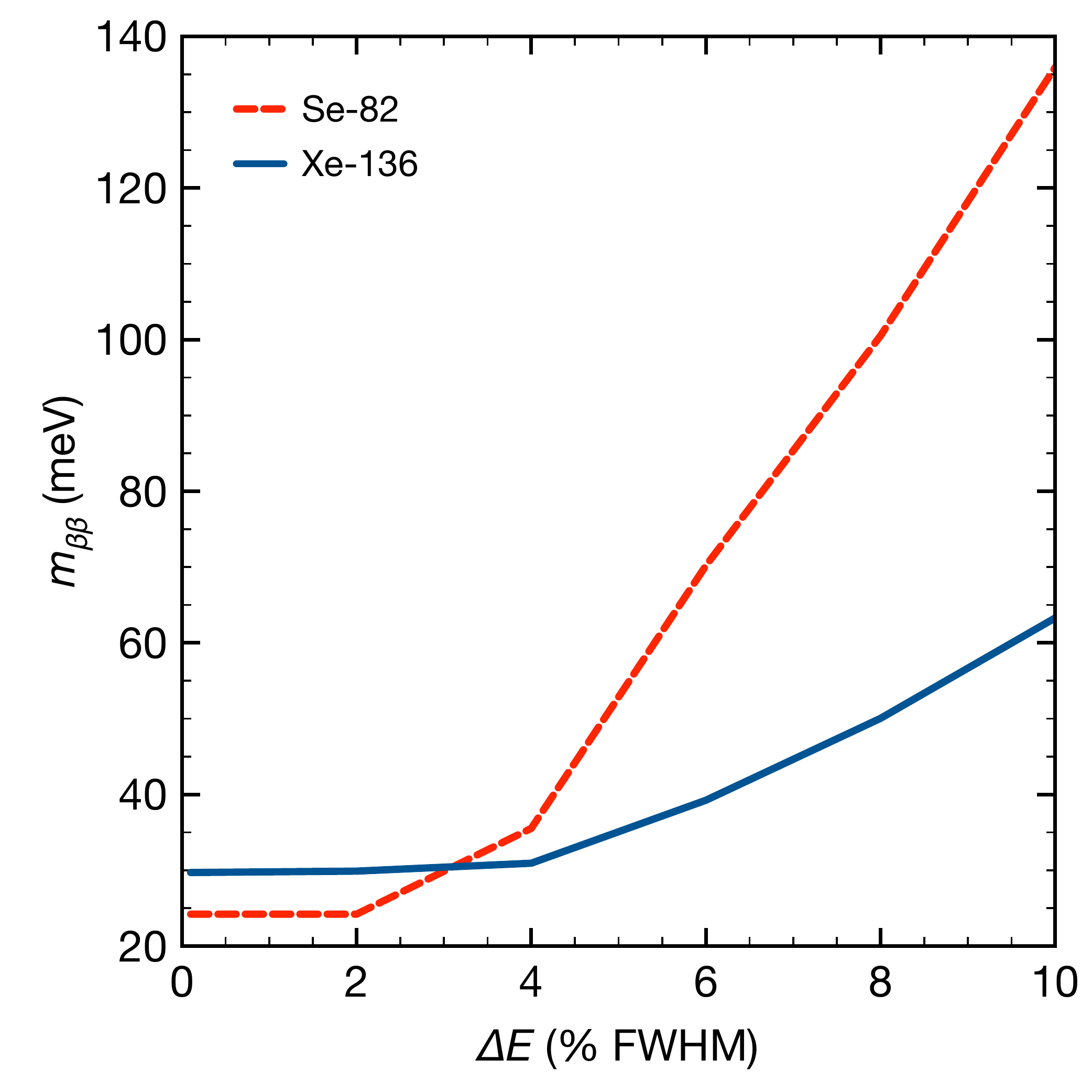}
\caption{Dependence on the energy resolution of the sensitivity to \mbb\ (at 90\% CL) of experiments with 500~$\mathrm{kg}\cdot\mathrm{year}$ of exposure and affected only by \bbtnu-decay backgrounds.} \label{fig:SensitivityVsResolution}
\end{figure}
%%%%%%%%%%

Experimental techniques with modest energy resolution (say, more than 3--4\% FWHM at the $Q$ value) are also interested in choosing an isotope with a relatively slow two-neutrino decay mode. The significance of the \bbtnu\ spectrum as a background depends on its spectral form and rate. It is irrelevant if the energy resolution of the experiment is effectively perfect, but as the energy resolution degrades, it can become a serious background. This is illustrated in Figure~\ref{fig:SensitivityVsResolution}, where the \mbb\ sensitivity (90\% CL) as a function of the energy resolution is shown for experiments using \XE\ and \SE. The experiments are assumed to have perfect detection efficiency and be affected only by \bbtnu\ backgrounds. In these idealized conditions, it is clear that \XE\ is preferable to \SE\ for low-resolution experiments thanks to its much longer \bbtnu-decay half-life \cite{Barabash:2015eza}.

The general conclusion is that there is no \emph{magic} candidate, no specially favoured or disfavoured isotope among those with a $Q$ value larger than 2~MeV. The choice of the \bb\ isotope is, therefore, ultimately driven by its procurement cost  and by the experimental technique.

%%%%%%%%%%%%%%%%%%%%%%%%%%%%%%%%%%%%%%%%
\subsection{Energy resolution} \label{subsec:EnergyResolution}
\begin{figure}
\centering
\includegraphics[width=0.45\textwidth]{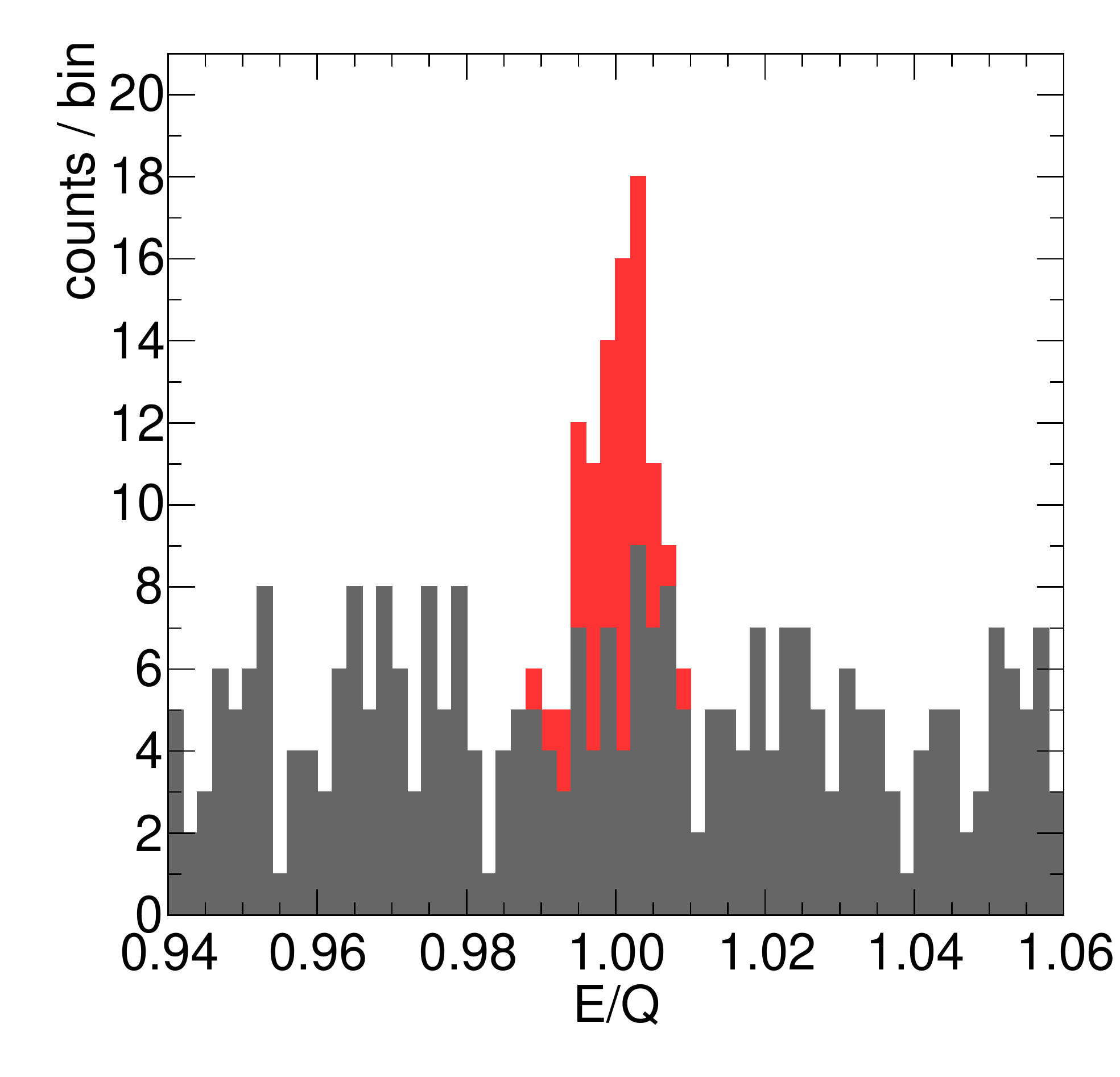}\\
\includegraphics[width=0.45\textwidth]{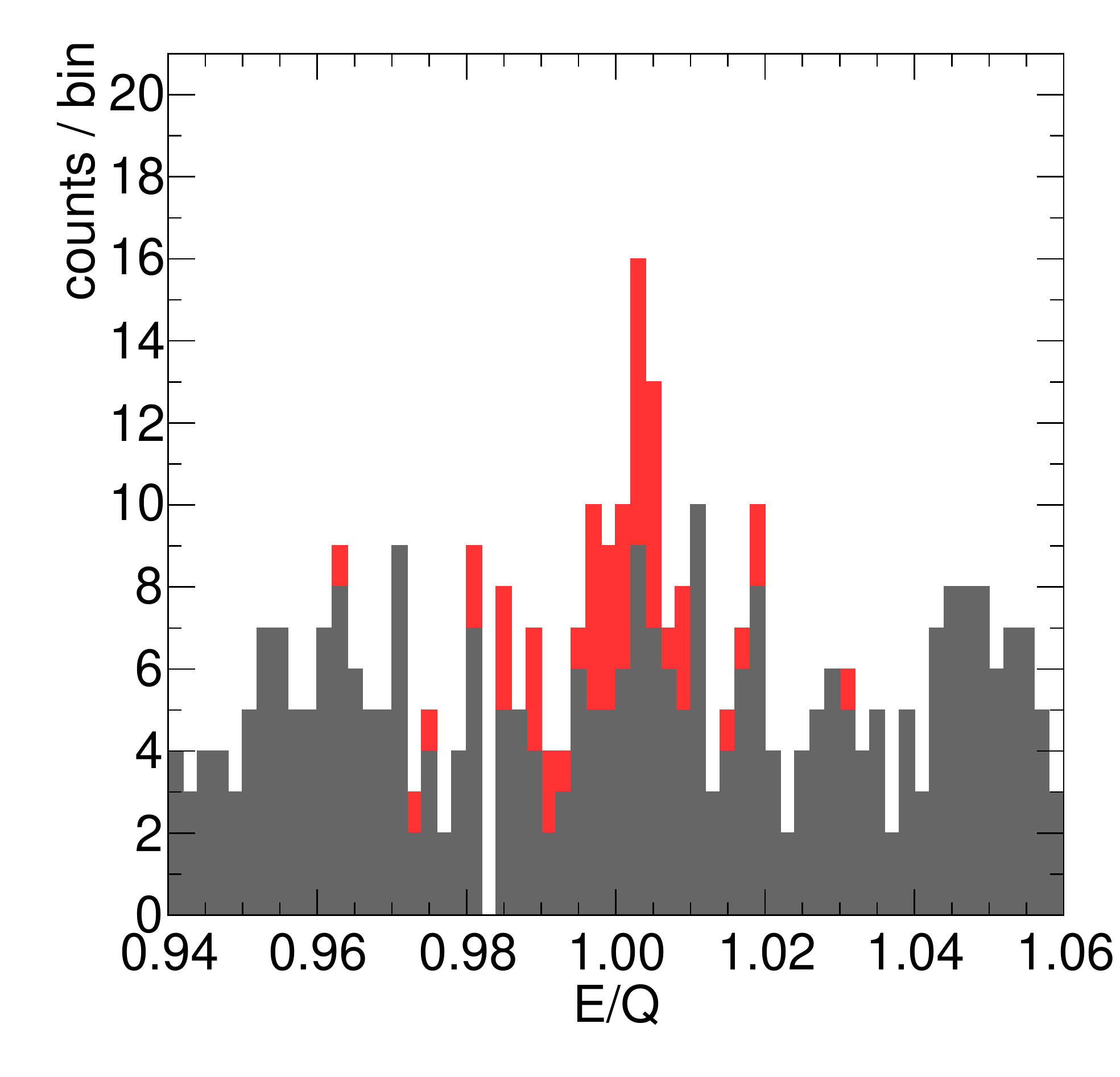}\\
\includegraphics[width=0.45\textwidth]{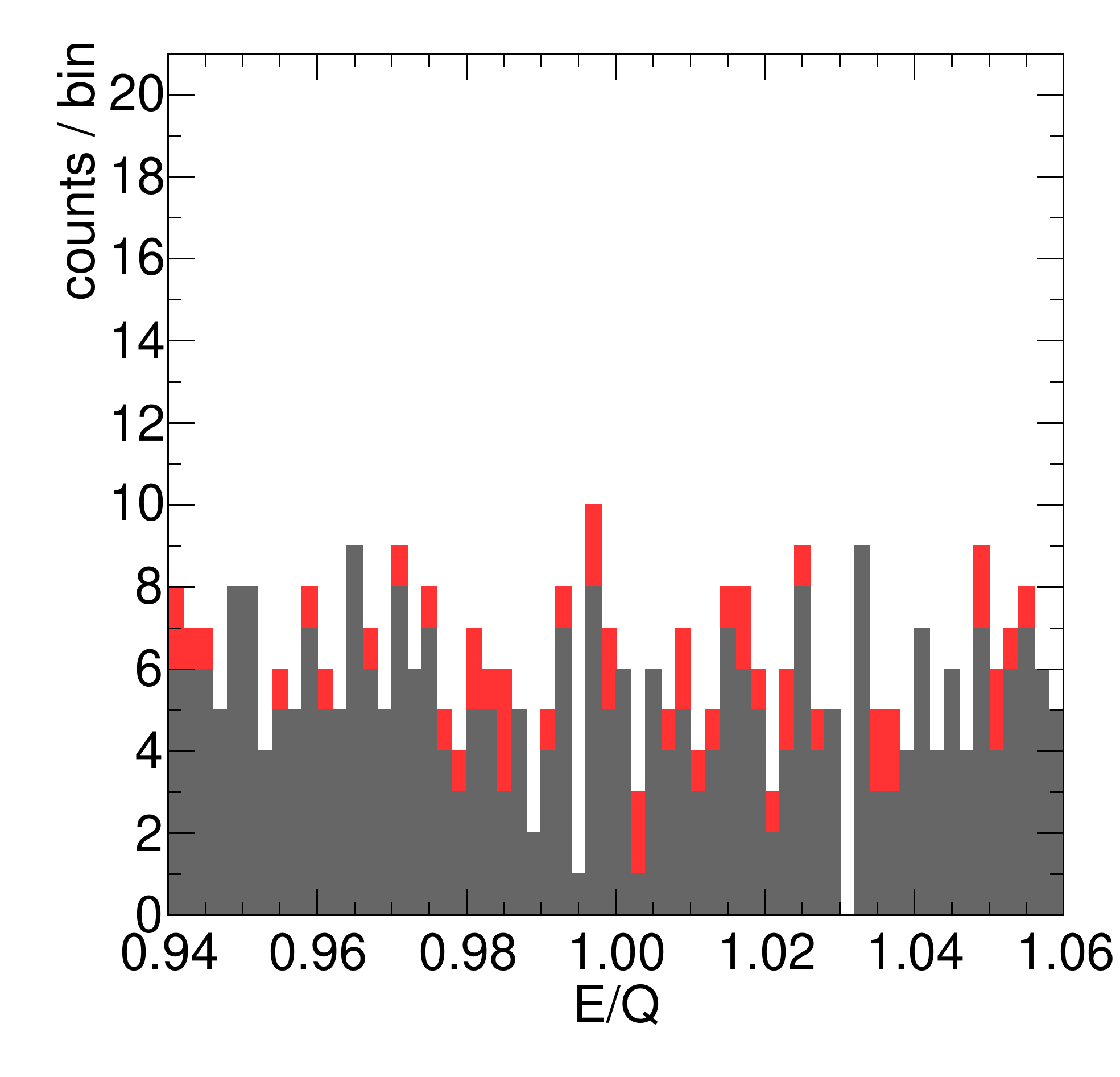}
\caption{Signal and background (red and grey stacked histograms, respectively) in the region of interest around \Qbb\ for three Monte Carlo experiments with the same signal strength (50 counts) and background rate (1 count/keV), but different energy resolution (top: 1\% FWHM; centre: 3.5\% FWHM; bottom: 10\% FWHM). The signal is distributed normally around \Qbb, while the background is assumed flat.} \label{fig:EnergyResolutionSNR}
\end{figure}

%%%
High energy resolution is a necessary condition (but not sufficient) for an ultimate \bbonu-decay experiment: it is the only protection against the intrinsic \bbtnu\ background (see Fig.~\ref{fig:SensitivityVsResolution}), and improves the signal-to-noise ratio in the region of interest around \Qbb. Figure~\ref{fig:EnergyResolutionSNR} illustrates the latter point: the energy region of interest around the $Q$ value is represented for three Monte Carlo experiments with the same signal and background, but different energy resolution. The signal is distributed normally around \Qbb, whereas the background is assumed flat in the window. The signal strength (50 counts) and the background rate (1~count/keV) correspond to typical values for a tonne-scale experiment and a Majorana neutrino mass of the order of 100~meV. A clear peak rising above the background is visible in the case of the experiment with better energy resolution (1\% FWHM). However, the peak is hardly discernible for the experiment with a 3\% FWHM resolution, and it has disappeared completely for the worst-resolution case (10\% FWHM). In conclusion, the better the energy resolution, the greater the \emph{discovery potential} of an experiment given certain conditions of signal and background. Therefore, the experimental techniques with modest or poor energy resolution have to compensate this deficiency reaching lower background rates and higher exposures.

%%%%%%%%%%
%%%%%%%%%%

%%%%%%%%%%%%%%%%%%%%%%%%%%%%%%%%%%%%%%%%
\subsection{Low background} \label{subsec:Backgrounds}
%%%
We have seen already that the  presence of background in the region of interest around \Qbb\ changes the regime of the sensitivity to \mbb\ of an experiment from an inverse square-root dependence on the exposure to an inverse fourth root. For this reason, the main developmental challenges for any double beta decay experiment are all concerned with the suppression of backgrounds. 

\begin{figure}
\centering
\includegraphics[width=0.5\textwidth]{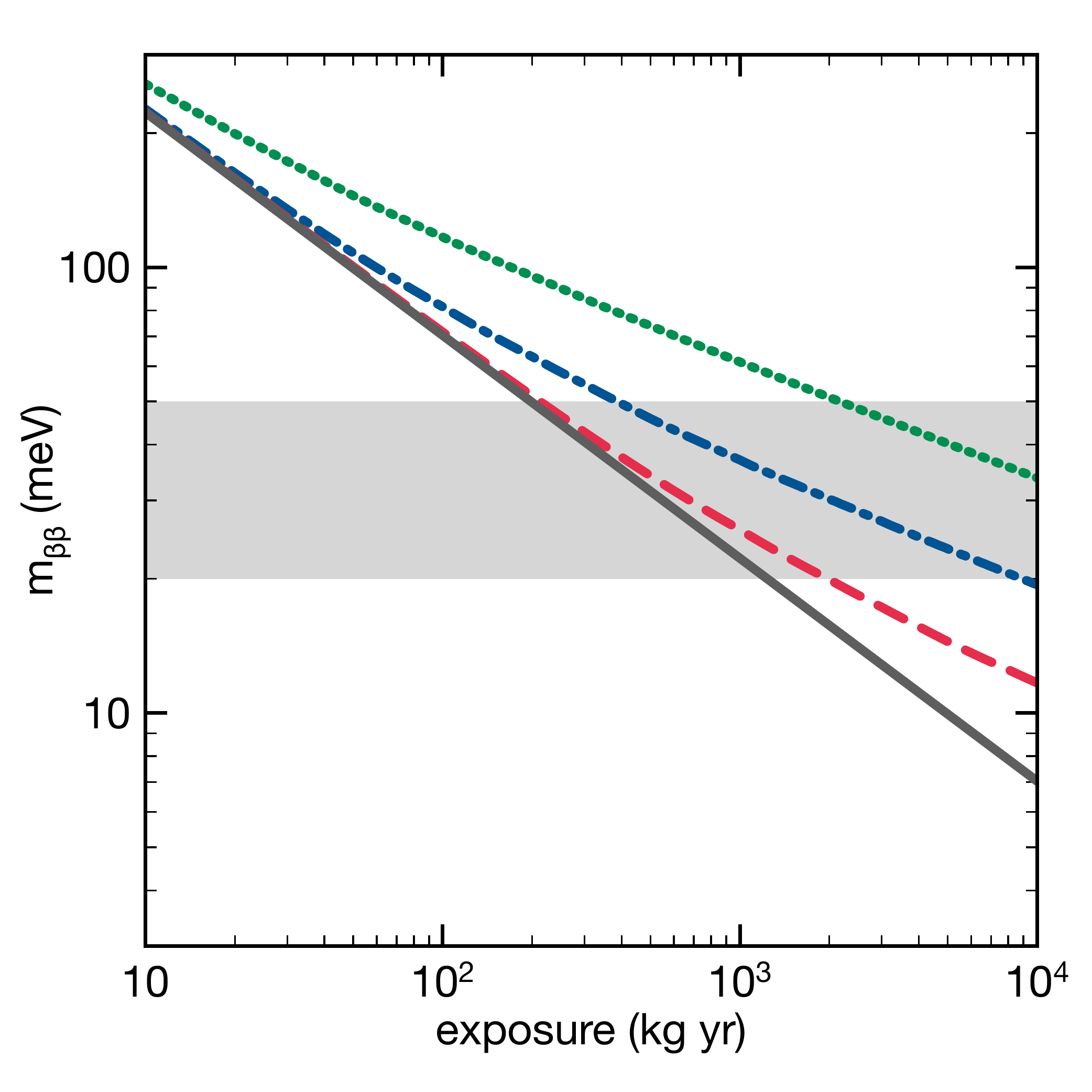}
\caption{Sensitivity to \mbb\ (at 90\% CL) of a \XE-based experiment with perfect signal detection efficiency and four different assumptions for the background rate within the energy region of interest around \Qbb: 0.1~$\mathrm{cts}~\mathrm{kg}^{-1}~\mathrm{yr}^{-1}$ (green, dotted line); 0.01~$\mathrm{cts}~\mathrm{kg}^{-1}~\mathrm{yr}^{-1}$ (blue, dash-dot line); 0.001~$\mathrm{cts}~\mathrm{kg}^{-1}~\mathrm{yr}^{-1}$ (red, dashed line) and background-free (grey, solid line). The IBM-2 nuclear matrix element \cite{Barea:2013bz} has been used to convert the half-life limits to \mbb. The grey band represents the inverted-hierarchy region of neutrino masses.} \label{fig:SensitivityXeVsBackground}
\end{figure}

Figure~\ref{fig:SensitivityXeVsBackground} shows the sensitivity (at 90\% CL) of a \XE-based\footnote{The conclusions reached here would be the same regardless of the \bb\ isotope used for the discussion.} experiment with four different assumptions for the background rate within the energy region of interest: 10$^{-1}$, 10$^{-2}$ and 10$^{-3}$ counts~kg$^{-1}$~yr$^{-1}$, and zero background. EXO-200 and KamLAND-Zen have achieved a background rate of approximately 0.19 and 0.04~cts~kg$^{-1}$~yr$^{-1}$, respectively. Both experiments have accumulated about 100~kg~yr of exposure, and thus, according to the figure, their sensitivities are slightly above 100~meV. In order to probe Majorana neutrino masses down to 20~meV, the tonne-scale versions of these experiment must improve the quoted background rates in more than one order of magnitude, reaching values close to 10$^{-3}$~counts~kg$^{-1}$~yr$^{-1}$. The exploration of the normal-hierarchy region of neutrino masses ($\mbb<20$~meV) will only be possible with huge, background-free experiments.

%%%%%%%%%%
%%%%%%%%%%

The natural radioactivity of detector components is usually the main background in \bbonu-decay experiments. Even though the half-lives of the natural decay chains are comparable to the age of the universe, they are very short compared to the half-life sensitivity of the new-generation experiments. Consequently, even traces of these nuclides can become a significant background. Particularly pernicious are \TL\ and \BI\  decay products of the thorium and uranium series, respectively  due to the high energy of the particles emitted in their decays. These isotopes are present at some level in all materials. Therefore, careful selection of \emph{radiopure} materials is mandatory for all \bbonu-decay experiments. New-generation detectors are being fabricated from components with activities as low as a few microbecquerels per kilogram or less.

Radon, another intermediate decay product of the uranium and thorium series, is also a concern for most experiments. It is one of the densest substances that remains a gas under normal conditions, and it is also the only gas in the atmosphere that only has radioactive isotopes. While the average rate of production of $^{220}$Rn (from the thorium decay series) is about the same as $^{222}$Rn, the longer half-life of the latter (3.8 days versus 55 seconds) makes it much more abundant. Being a noble gas, radon is chemically not very reactive and can diffuse easily through many materials, infiltrating into the active region of the detectors. Radon progenies, also radioactive, tend to be charged and adhere to surfaces or dust particles.  The impact of radon can be mitigated by flushing the detector surroundings with pure nitrogen or by installing radon traps in the laboratory air circulation systems. 

In addition to the backgrounds coming from radioactive impurities in detector components, there are external backgrounds originating outside the detector. These can be suppressed by placing the detector underground and by enclosing it in a shielding system. Very efficient shielding and additional detection signatures such as track reconstruction can compensate the benefits of a very deep location. Several underground facilities are currently available to host physics experiments around the world \cite{Bettini:2011zza}. 
At the depths of underground laboratories, muons and neutrinos are the only surviving radiation from the atmosphere and outer space. Future very massive detectors will have to deal with the irreducible external background due to elastic electron scattering of solar neutrinos \cite{deBarros:2011qq}. Muon interactions can produce high-energy secondaries such as neutrons or electromagnetic showers. Charged backgrounds can be easily eliminated using a veto system. Neutrons, on the other hand, are a more serious problem. They can have sizable penetrating power, impinging on the detector materials and \emph{activating} them, ultimately resulting in radioactive nuclides. Detectors can be shielded against neutrons with layers of hydrogenous material. Cosmogenic activation is, of course, more severe on surface . Therefore, for experiments using materials that can get activated (like germanium or copper) \cite{Cebrian:2010zz}, underground fabrication and storage of the detector components may be essential. 

Natural radioactivity in the rock of the underground caverns results in a gamma-ray flux that can interact in the detector producing background. Dense, radiopure materials such as lead or copper are used as shielding to attenuate this background. Water, being inexpensive and easy to purify, is also a good alternative for shielding against $\gamma$ rays.

Besides the passive background reduction techniques mentioned above, most experiments use now \emph{active} methods for the discrimination of signal and background: reconstruction of the event topology, pulse-shape discrimination, combination of detection signatures, etc. A unique possibility offered by xenon-based experiments is that all backgrounds except the two-neutrino decay mode could be effectively removed by identification of the daughter barium ion using atomic laser resonant spectroscopy \cite{Moe:1991ik}.

%%%%%%%%%%%%%%%%%%%%%%%%%%%%%%%%%%%%%%%%
\subsection{Detection efficiency} \label{subsec:Efficiency}
%%%
Neutrinoless double beta decay is extremely rare, if existent at all. A high detection efficiency is, therefore, an important requirement for a \bb\ experiment, as clearly stated by Eq.~(\ref{eq:Sensitivity3}). To obtain the same increase in \mbb\ sensitivity attained by doubling the efficiency, the mass would have to be increased by a factor of 4, assuming the same background. In general, the simpler the detection scheme, the higher the detection efficiency. For instance, pure calorimetric approaches such as germanium diodes or bolometers have detection efficiencies in excess of 80\%. This is to be contrasted with experiments performing, for example, particle tracking, which will typically result in significant efficiency loss. Homogeneous detectors, where the source material is the detection medium, provide in principle higher efficiency than the separate-source approach for a number of reasons, including geometric acceptance or absorption in the \bb\ source. That being said, some homogeneous detectors may use part of the mass close to the detector boundaries for self-shielding against external backgrounds, paying it with efficiency loss.

%%%%%%%%%%%%%%%%%%%%%%%%%%%%%%%%%%%%%%%%
\subsection{Exposure} \label{subsec:Exposure}
%%%
Thousands of kilograms of \bb\ source will be needed to explore the extremely long \bbonu\ half-lives corresponding to the inverted hierarchy of neutrino masses. Most collaborations searching for \bbonu\ decay are advertising already future tonne-scale versions of their experiments. However, not all the technologies are equally suitable for that purpose. The scalability of each experimental technique will be, therefore, one of the key points --- together with the detector performance at the 100-kg stage --- for the evaluation of these proposals.

Large-scale production of the \bb\ isotopes will represent a technical and logistic challenge, as well as a significant fraction of the total cost of the detectors. Some of the most popular \bb\ isotopes are, in fact, quite rare in Earth. For example, the annual world production of germanium or tellurium is a few hundred tonnes, and much less, a few tens of tons, in the case of xenon \cite{Biller:2013wua}. Moreover, with the exception of \TE, which represents approximately one third of all tellurium, the isotopic abundance of the \bb-decaying isotopes is around or below 10\%, requiring isotopic enrichment in order to obtain large, concentrated masses. In fact, this has been so far the driving cost in the procurement of the isotopes. The most cost-effective enrichment technology is centrifugal separation \cite{ Giuliani:2012zu, Biller:2013wua, Tikhomirov:2000td}, but it is only possible for elements with a stable gas compound. Affordable enrichment of large quantities of those species with no gas compound, such as \CA\ or \ND, is not possible at present. Centrifugation of xenon, being a noble gas, is, of course, simpler (and hence cheaper) than that of metalloids such as germanium. Therefore, from this point of view, \XE\ would be a particularly favourable isotope to use for a tonne-scale experiment.

%%%%%%%%%%%%%%%%%%%%%%%%%%%%%%%%%%%%%%%%%%%%%%%%%%%%%%%%%%%%
\section{Current generation of experiments} \label{sec:CurrentExperiments}
%%%
Three experiments of the current generation --- EXO-200, KamLAND-Zen and GERDA --- have been operating for a few years already, and at least five other --- CUORE, {\scshape Majorana}, NEXT, SNO+ and the SuperNEMO demonstrator --- plan to start taking data soon. In this section we describe them succinctly, highlighting the main features of each technique. 

%%%%%%%%%%%%%%%%%%%%%%%%%%%%%%%%%%%%%%%%
\subsection{CUORE} \label{subsec:CUORE}
%%%
The \emph{Cryogenic Underground Observatory for Rare Events} (CUORE) will search for the \bbonu\ decay of \TE\ using TeO$_{2}$ crystal bolometers \cite{Artusa:2014lgv}. When these crystals are cooled to 10~mK, their heat capacity becomes so small that the energy deposited by interacting particles is measurable as a rise in temperature. The crystals, therefore, function as highly sensitive calorimeters. This technique was used for the first time in \bbonu-decay searches by the MiDBD \cite{Arnaboldi:2002te} and Cuoricino \cite{Andreotti:2010vj} $^{130}$Te experiments. 

%%%%%%%%%%
\begin{figure}
\centering
\includegraphics[scale=.225]{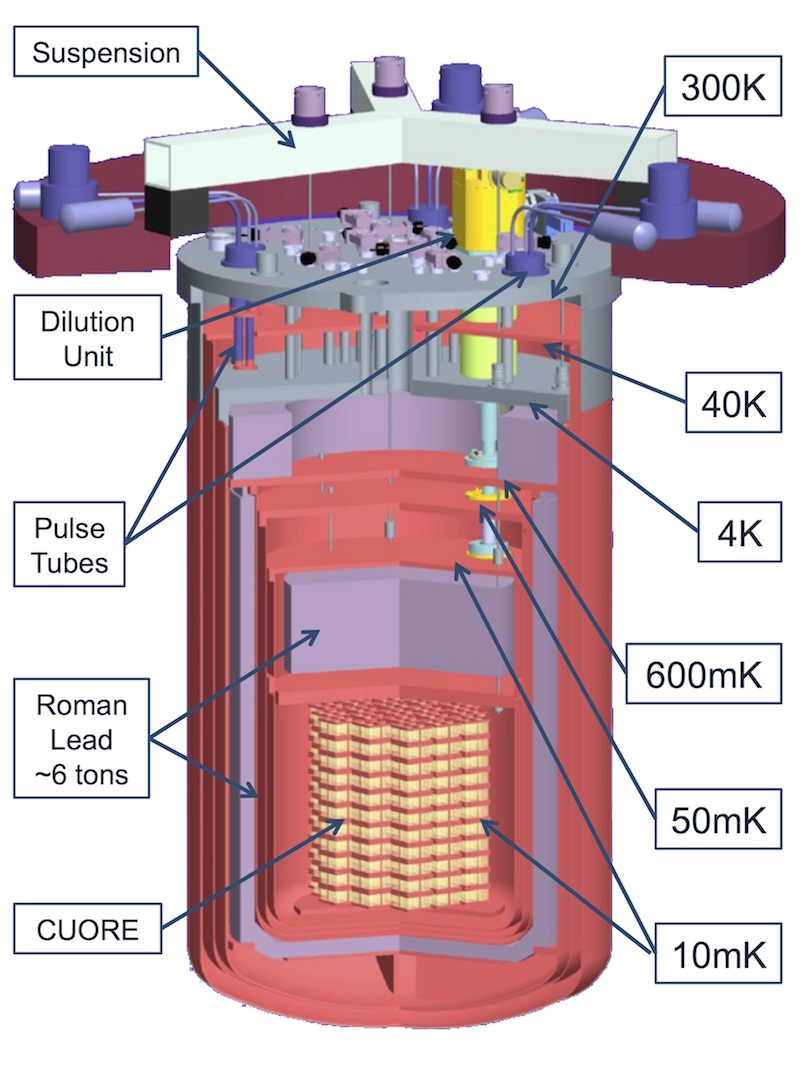}
\caption{Cutaway view of the CUORE bolometers inside the cryostat, consisting of six nested copper vessels at 300~K (outer vacuum chamber), 40~K, 4~K (inner vacuum chamber), 0.6~K (still), 0.05~K (heat exchanger), and 0.01~K (mixing chamber). Several layers of radiopure lead shield the bolometers from external radiation. Reproduced from Artusa et al.\ (2014) \cite{Artusa:2014lgv}.} \label{fig:Cuore}
\end{figure}
%%%%%%%%%%

CUORE, currently under construction at the Laboratori Nazionali del Gran Sasso, will consist of 988 bolometers arranged in 19 vertical towers held by a copper frame. The basic detector element is a $5\times5\times5$~cm$^{3}$ TeO$_{2}$ crystal of 750~g instrumented with a temperature sensor and a resistive heater. The total mass of the bolometers will be 741 kg, of which 206~kg are $^{130}$Te. The bolometer towers will be housed in a cryostat composed of six nested copper vessels (see Figure~\ref{fig:Cuore}). Two cold lead shields will shield the bolometers from radiation originating in the cryostat: a layer of ancient Roman lead 6~cm thick located between the two middle copper vessels will shield the detectors from radioactivity in the outer vessels, and a disc 31 cm thick made of modern and Roman lead and located below the mixing chamber plate will shield the detectors from radioactivity in the overhead apparatus. The cryostat will be surrounded by a 73-tonnes octagonal external shield designed to screen the detector from environmental $\gamma$ rays and neutrons. The shield has three layers: an outermost layer 20~cm thick consisting of a floor and sidewalls of polyethylene to thermalize and absorb neutrons; a side layer 2~cm thick of boric-acid powder to absorb neutrons; and an innermost layer of lead bricks of at least 25~cm thickness to absorb $\gamma$ rays.

CUORE aims at improving the sensitivity of Cuoricino, its predecessor, by more than a factor of 30 by operating a larger, cleaner, better-shielded detector with enhanced energy resolution. The expected energy resolution (FWHM) of the CUORE crystals is 5~keV at the $Q$ value of \TE\ (2528~keV) \cite{Artusa:2014lgv}. This resolution has already been achieved in tests performed during the R\&D phase. In Cuoricino, the average background rate in the region of interest was 0.58~\ckky. Three main contributions were identified \cite{Alessandria:2011rc}: ($30\pm10$)\% of the measured  background in the region of interest was due to multi-Compton events due to the 2615-keV gamma ray from the thorium decay chain from the contamination of the cryostat shields; ($10\pm5$)\% was due to surface contamination of the crystals (primarily degraded alphas from the natural decay chains); and ($50\pm20$)\% is ascribed to similar surface contamination of inert materials surrounding the crystals, most likely copper. On the basis of this result, the R\&D for CUORE has pursued two major complementary lines: the reduction of surface contamination and the selection of extremely radiopure construction materials. The goal is achieving a background rate in the region of interest of 0.03--0.04~\ckky\ \cite{Alessandria:2011rc}.

CUORE-0, a single tower of CUORE containing 52 TeO$_{2}$ crystals, is in operation inside the Cuoricino cryostat since March 2013. It will directly test the level of backgrounds of the CUORE experiment and improve the Cuoricino sensitivity to the \TE\ \bbonu-decay half-life. CUORE is now in an advanced state of construction. The Collaboration plans to complete the integration and commissioning of the detector at the end of 2014, and commence data taking in the first half of 2015 \cite{Artusa:2014lgv}.

%%%%%%%%%%%%%%%%%%%%%%%%%%%%%%%%%%%%%%%%
\subsection{EXO} \label{subsec:EXO}
%%%
The \emph{Enriched Xenon Observatory} (EXO) is an experimental program searching for neutrinoless double beta decay using \XE. The first phase of the experiment, EXO-200 \cite{Auger:2012gs}, consists in a 200-kg liquid xenon (LXe) time projection chamber that has been taking physics data at the Waste Isolation Pilot Plant (WIPP), in New Mexico, USA, since early May 2011. The results produced so far by the experiment include the first observation of the \bbtnu\ decay of \XE\ \cite{Albert:2013gpz} and some of the most stringent limits so far on the effective Majorana neutrino mass \cite{Albert:2014awa, Auger:2012ar}. Building on the success of EXO-200, the EXO Collaboration has started the R\&D work for a future multi-tonne LXe experiment called nEXO.  

%%%%%%%%%%
\begin{figure}
\centering
\includegraphics[width=\textwidth]{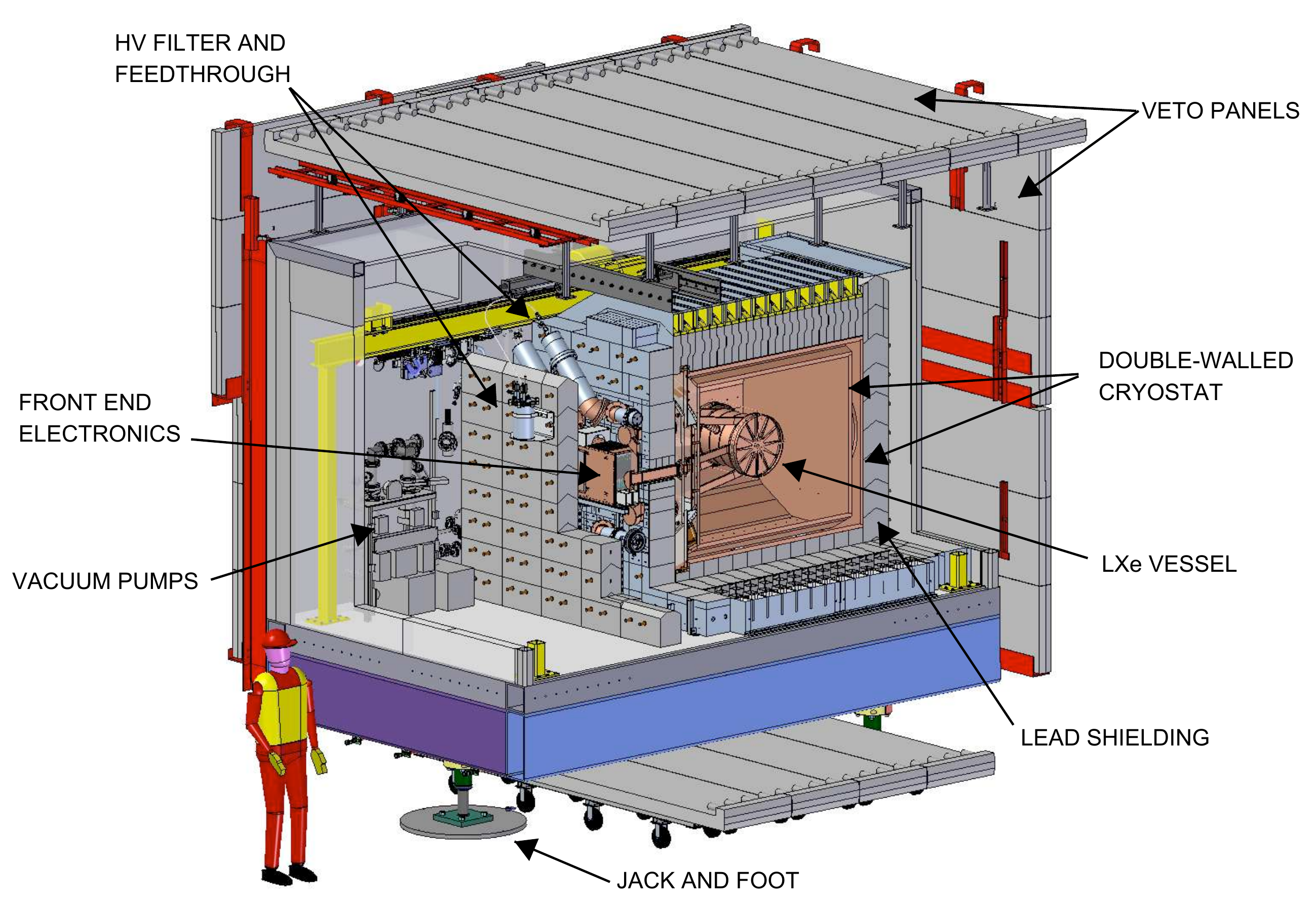}
\caption{Drawing of the EXO-200 detector, cryostat and shielding. Reproduced from Auger et al.\ (2012) \cite{Auger:2012gs}.} \label{fig:EXO200}
\end{figure}
%%%%%%%%%%

The EXO-200 detector is a cylindrical TPC, about 40~cm in diameter and 44~cm in length, with two drift regions separated in the centre by a transparent cathode. The TPC measures the 3D coordinates and energy of ionization deposits in the LXe by simultaneously collecting the scintillation light and the charge. Charge deposits spatially separated by about 1~cm or more are individually observed with a position accuracy of a few millimetres. A pair of crossed wire planes collects the ionization charge and measures its amplitude and transverse coordinates, and arrays of avalanche photodiodes (APDs) located behind the wire planes measure the scintillation light. The sides of the chamber are covered with teflon sheets that act as VUV reflectors improving the light collection. The xenon, enriched to 80.6\% in \XE, is held inside a thin copper vessel immersed in a cryofluid that also shields the detector from external radioactive backgrounds. The cryofluid is maintained at $\sim$167 K inside a vacuum-insulated copper cryostat. Further shielding is provided by at least 25~cm of lead in all directions. The entire assembly is housed in a clean-room located underground at WIPP. Four of the six sides of the clean-room are instrumented with plastic scintillator panels recording the passage of cosmic ray muons. Figure~\ref{fig:EXO200} shows the overall detector and shielding arrangement.

The fiducial volume selected in the data analysis contains a \XE\ mass of 76.5~kg. Signal detection efficiency is estimated to be 84.6\%. Thanks to the simultaneous measurement of the ionization and scintillation signals, EXO-200 reaches an energy resolution of 3.6\% FWHM at the $Q$ value of \XE. The last published analysis \cite{Albert:2014awa}, analyzes an exposure of 100~kg~yr (736~mol~yr). The estimate of the background in a $\pm2\sigma$ window around \Qbb\ is $31.1\pm1.8\,(\mathrm{stat})\pm3.3\,(\mathrm{syst})$ counts, or $(2.1\pm0.3)\times10^{-3}$~\ckky. The dominant backgrounds arise from the thorium series (16.0 counts), the uranium series (8.1 counts) and $^{137}$Xe (7.0 counts). This amount of $^{137}$Xe is consistent with estimates from studies of the activation of \XE\ in muon-veto-tagged data. The EXO Collaboration reports a 90\% CL lower limit on the half-life of \XE\ of $1.1\times10^{25}$~yr. This corresponds to an upper limit on the Majorana neutrino mass of 190--450 meV.

The EXO Collaboration is planning a new, next-generation detector, nEXO, that would use 5~tonnes of enriched xenon \cite{Albert:2014afa}. Many of the detector concepts and implementation in EXO-200 are, in principle, scalable to a larger mass of xenon, and the self-shielding improves with larger mass. The detector would be placed in a large water shield instead of the lead shield used for EXO-200 and a deeper site would be chosen to reduce the cosmogenic neutron backgrounds.

%%%%%%%%%%%%%%%%%%%%%%%%%%%%%%%%%%%%%%%%
\subsection{GERDA and {\scshape Majorana}} \label{subsec:GERDA}
\begin{figure}
\centering
\includegraphics[width=.75\textwidth]{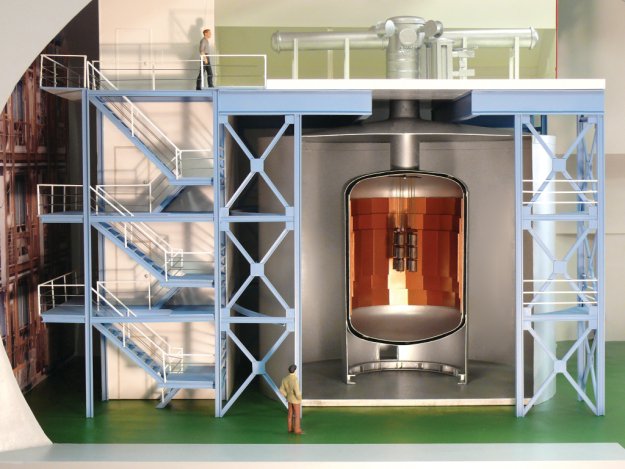}
\caption{Artist's view (HPGe strings not to scale) of the GERDA detector at LNGS. Reproduced from Macolino (2013) \cite{Macolino:2013hca}.} \label{fig:GERDA}
\end{figure}

%%%
The \emph{GERmanium Detector Array} (GERDA) experiment, located in Hall A of the Laboratori Nazionali del Gran Sasso (LNGS), is searching for the \bbonu\ decay of \GE\ using bare high-purity germanium (HPGe) diodes immersed in a cryogenic bath of liquid argon (LAr) \cite{Ackermann:2012xja}. The HPGe detectors are arranged in strings and mounted in special low-mass holders made of ultra-pure copper and PTFE. The strings are suspended inside a vacuum-insulated stainless steel cryostat of 4.2~m diameter and 8.9~m height filled with LAr. A copper lining 6~cm thick covers the inner cylindrical shell of the cryostat. The cryostat is placed in a 590-m$^{3}$ water tank instrumented with PMTs which serves as a Cherenkov muon veto as well as a gamma and neutron shield. A drawing of the entire system is shown in Figure~\ref{fig:GERDA}.

%%%%%%%%%%
%%%%%%%%%%

The GERDA experiment was planned in two physics stages.
Phase I, which started in November 2011 and ended in March 2013, has used refurbished semi-coaxial HPGe detectors from the Heidelberg-Moscow and IGEX experiments, which are isotopically enriched to 86\% in \GE, plus a non-enriched detector from the GENIUS-TF project, totalling a mass of 17.67~kg. In addition to these detectors, 5 broad-energy germanium (BEGe) diodes foreseen for the second phase of the experiment were deployed in July 2012. The exposure-averaged energy resolution (FWHM) of the detectors at the $Q$ value of \GE\ is $(4.8\pm0.2)$~keV for the semi-coaxial detectors and $(3.2\pm0.2)$~keV for the BEGe detectors.

The GERDA Collaboration has published a measurement of the \bbtnu\ half-life of \GE, $T^{2\nu}_{1/2} = 1.84^{+0.09}_{-0.08}\,(\mathrm{stat})\, ^{+0.11}_{-0.06}\,(\mathrm{syst})\times10^{21}$~years \cite{Agostini:2012nm}, and a limit to the \bbonu\ half-life, $T^{0\nu}_{1/2}(\GE)>2.1\times10^{25}$~yr (90\% CL) \cite{Agostini:2013mzu}. This analysis was done with 17.9~kg~yr of exposure. The achieved background rate was $(11\pm2)\times10^{-3}$~\ckky,  and the \bbonu\ signal acceptance was 86\%. The GERDA result is consistent with the limits by
Heidelberg-Moscow and IGEX. The combination of the results of the three experiments yields a limit of $3.0\times10^{25}$~years (90\% CL) \cite{Agostini:2013mzu}. Therefore, the long-standing claim for a \bbonu\ signal in \GE\ is strongly disfavoured.

In Phase II, besides the increase of the active mass by about 20 kg (30 BEGe detectors), the main goal is to further reduce the background by one order of magnitude thanks to several improvements in the detector setup (instrumentation of the LAr bath, materials of higher radiopurity in the vicinity of the detectors, etc.) \cite{Macolino:2014vya}. 

For the very long term, it is foreseen a third phase of the experiment with about 1 tonne of $^{76}$Ge together with further reduction of background. Such an effort would be feasible only in a word-wide collaboration with the {\scshape Majorana} project \cite{Abgrall:2013rze}, which is following a more classic approach than GERDA in the design of a germanium-based experiment. The {\scshape Majorana} Collaboration is building a modular setup composed of two cryostats built from ultra-pure electroformed copper, with each cryostat capable of housing over 20 kg of HPGe detectors. The baseline plan calls for 30 kg of the detectors to be built from Ge material enriched to 86\% in isotope 76 and 10 kg fabricated from natural Ge. Starting from the innermost cavity, the cryostats will be surrounded by an inner layer of electroformed copper, an outer layer of oxygen-free copper, high-purity lead, an active muon veto, polyethylene, and borated polyethylene. The cryostats, copper, and lead shielding will all be enclosed in a radon exclusion box. The entire experiment will be located in a clean room at the Sanford Underground Research Facility (SURF) in South Dakota, USA. The goal is to demonstrate a background rate of 3 counts per tonne and per year in the 4-keV wide region of interest. The detector should be in operation in 2015.

%%%%%%%%%%%%%%%%%%%%%%%%%%%%%%%%%%%%%%%%
\subsection{KamLAND-Zen}
\begin{figure}
\centering
\includegraphics[width=0.65\textwidth]{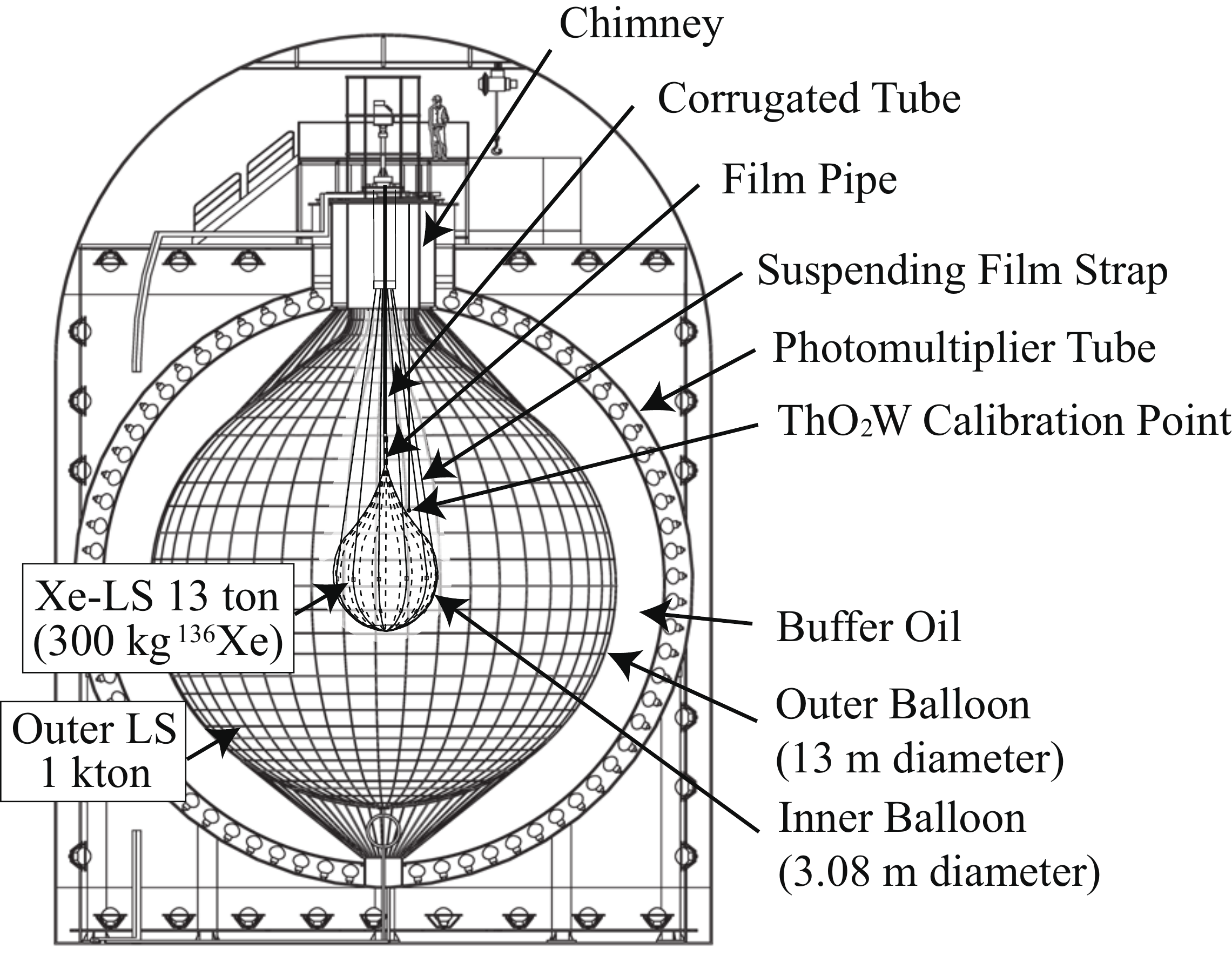}
\caption{Schematic drawing of the KamLAND-Zen detector. Reproduced from Gando et al.\ (2012) \cite{KamLANDZen:2012aa}.} \label{fig:KamLANDZen}
\end{figure}

%%%
The KamLAND-Zen experiment is searching for the \bbonu\ decay of \XE\ using enriched xenon dissolved in liquid scintillator, a technique first proposed by R.~Raghavan in 1994 \cite{Raghavan:1994qw}. The experiment reuses the neutrino KamLAND detector \cite{Abe:2009aa}, located at the Kamioka Observatory, Japan. The KamLAND-Zen detector, shown in Figure~\ref{fig:KamLANDZen}, is composed of two concentric transparent balloons. The inner one, 3.08~m diameter and fabricated from 25 $\mu$m thick nylon film, contains 13 tonnes of Xe-loaded liquid scintillator. The outer balloon, 13~m in diameter, contains 1~kilotonne of pure liquid scintillator, and serves as an active shield for external gamma background as well as a detector for internal radiation from the inner balloon. Buffer oil between the outer balloon and an 18~m diameter spherical stainless-steel containment tank shields the detector from external radiation. Scintillation light is recorded by 1325 17-in and 554 20-in photomultiplier tubes mounted on the stainless-steel tank, providing 34\% solid-angle coverage. The containment tank is surrounded by a 3.2-kt water-Cherenkov outer detector. The Xe-loaded scintillator consists of 82\% decane and 18\% pseudocumene by volume, 2.7 g/litre of the fluor PPO, and 2--3\% by weight of enriched xenon gas \cite{KamLANDZen:2012aa}, corresponding to approximately 350~kg of \XE.

%%%%%%%%%%
%%%%%%%%%%

KamLAND-Zen, which has been collecting physics data since late 2011, has published a measurement of the half-life of the \bbtnu\ decay of \XE, $2.38\pm0.02~(\mathrm{stat})\pm0.14~(\mathrm{syst})\times10^{21}$~years \cite{KamLANDZen:2012aa}, and a limit to the half-life of the \bbonu\ decay, $2.6\times10^{25}$~years (90\% CL)~\cite{Asakura:2014lma, Gando:2012zm}. The energy resolution of the detector is 9.9\% FWHM at the $Q$ value of \XE. The achieved background rate in the region of interest is approximately $1.4\times10^{-4}$~\ckky, thanks to a tight selection cut in the fiducial volume and the identification of $^{214}$Bi events via Bi-Po tagging \cite{Asakura:2014lma}.

%%%%%%%%%%%%%%%%%%%%%%%%%%%%%%%%%%%%%%%%
\subsection{NEXT}
%%%
The \emph{Neutrino Experiment with a Xenon TPC} (NEXT) \cite{Gomez-Cadenas:2013lta} will search for the neutrinoless double beta decay of \XE\ using a high-pressure xenon gas time projection chamber. Such a detector provides several valuable features: a) excellent \emph{energy resolution}, close, in principal, to 0.3\% FWHM at \Qbb; b) {tracking capabilities} that can be used for the distinction of signal (two electron emitted from a common vertex) and background (single electrons, mostly); c) a \emph{fully active and homogeneous} detector with no dead regions; and d) \emph{scalability} to large detector masses.

%%%%%%%%%%
\begin{figure}
\centering
\includegraphics[width=0.9\textwidth]{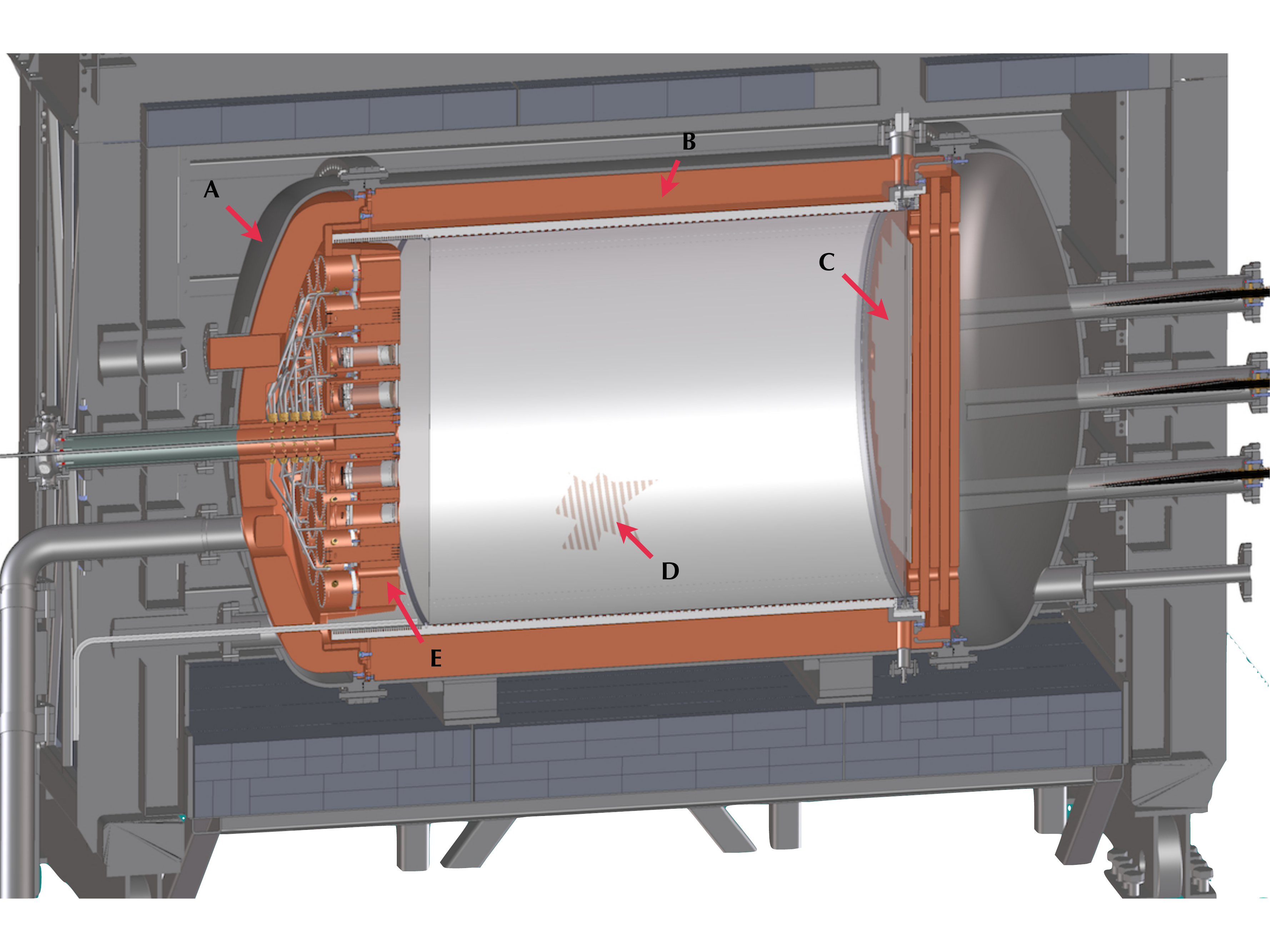}
\caption{A cross-section drawing of the NEXT-100 detector. The pressure vessel (A) is made of stainless steel Grade 316Ti, and its dimensions are 130~cm inner diameter, 222~cm length and 1~cm thick walls, for a total mass of 1\,200 kg. The inner copper shield (B) is made of ultra-pure copper bars and is 12~cm thick, with a total mass of 9\,000 kg. The time projection chamber (D) includes the field cage, cathode, EL grids and HV penetrators. The light tube is made of thin teflon sheets coated with TPB (a wavelength shifter). The energy plane (E) is made of 60 PMTs housed in copper enclosures. The tracking plane (C) is made of MPPCs arranged into 8$\times$8 boards.} \label{fig.NEXT100}
\end{figure}
%%%%%%%%%%

Following ideas introduced by D.\ Nygren (2009) \cite{Nygren:2009zz}, the design of the NEXT-100 detector (Figure \ref{fig.NEXT100}) is optimised for energy resolution by using proportional electroluminescent (EL) amplification of the ionisation signal. The detection process involves the use of the prompt scintillation light from the gas as start-of-event time, and the drift of the ionisation charge to the anode by means of an electric field ($\sim0.3$ kV/cm at 15 bar) where secondary EL scintillation is produced in the region defined by two highly transparent meshes, between which there is a field of $\sim20$ kV/cm at 15 bar. The detection of EL light provides an energy measurement using photomultipliers (PMTs) located behind the cathode (the \emph{energy plane}) as well as tracking through its detection a few mm away from production at the anode, via a dense array of silicon photomultipliers (the \emph{tracking plane}).

The R\&D phase of the experiment was carried out with the large-scale prototypes NEXT-DEMO and NEXT-DBDM. NEXT-DEMO is as a large-scale prototype of NEXT-100. The pressure vessel has a length of 60 cm and a diameter of 30 cm. The vessel can withstand a pressure of up to 15 bar and hosts typically 1--2 kg of xenon. NEXT-DEMO is  equipped with an energy plane made of 19 Hamamatsu R7378A PMTs and a tracking plane made of 256 Hamamatsu SiPMs. 

The detector has been operating successfully since 2011 and has demonstrated: (a) very good operational stability, with no leaks and very few sparks; (b) good energy resolution ; (c) track reconstruction with PMTs and with SiPMs coated with TPB; (d) excellent electron drift lifetime, of the order of 20 ms. Its construction, commissioning and operation has been instrumental in the development of the required knowledge to design and build the NEXT detector.

The NEXT-DBDM prototype is a smaller chamber, with only 8 cm drift, but an aspect ratio (ratio diameter to length) similar to that of NEXT-100. The device has been used to perform detailed energy resolution studies, as well as studies to characterise neutrons in an HPXe. NEXT-DBDM achieves a resolution of 1\% FWHM at 660 keV and 15 bar, which extrapolates to 0.5\% at \Qbb.

%%%%%
\begin{figure}
\centering
\includegraphics[width= 0.8\textwidth]{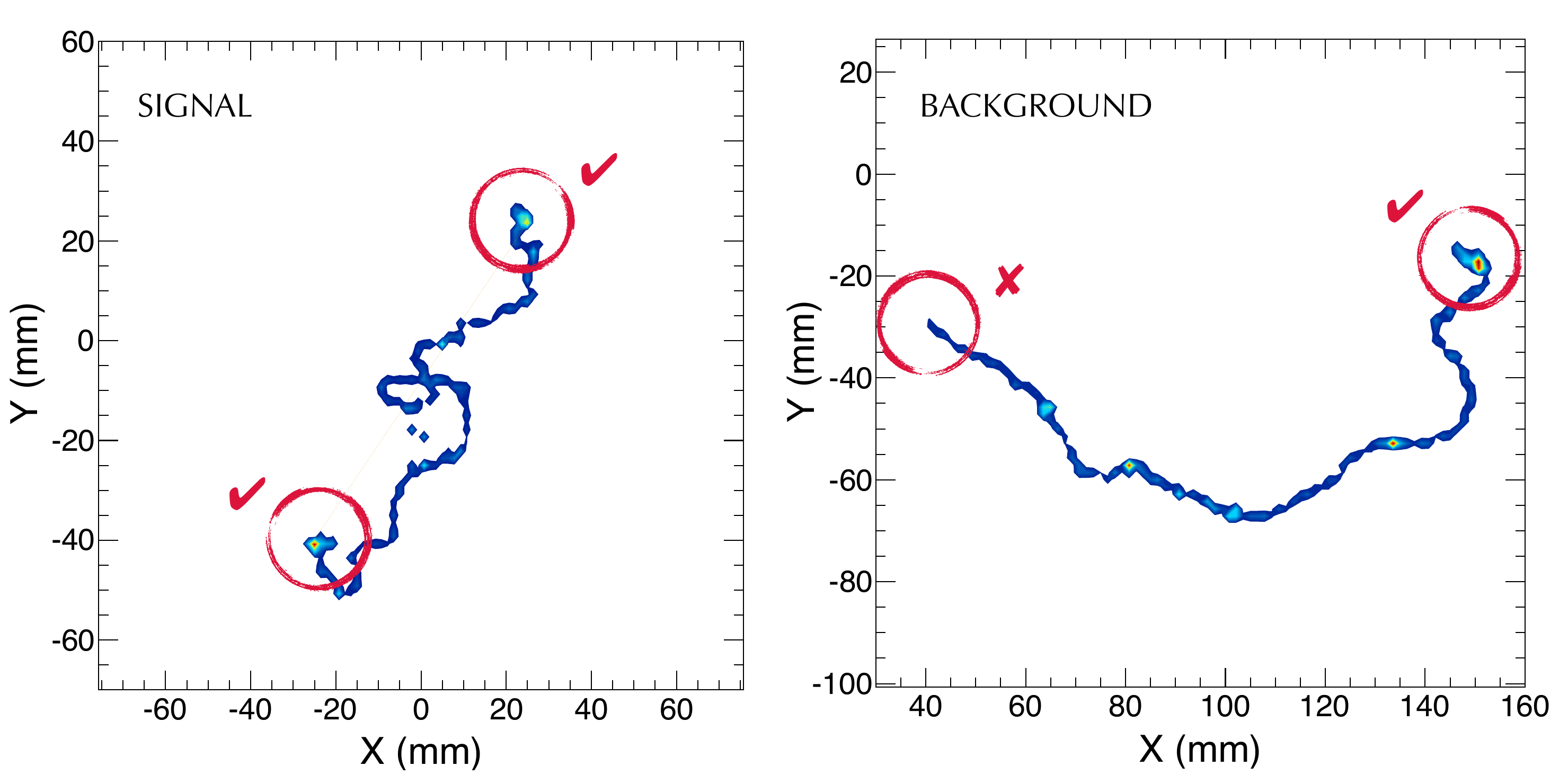}
\includegraphics[width=0.45\textwidth]{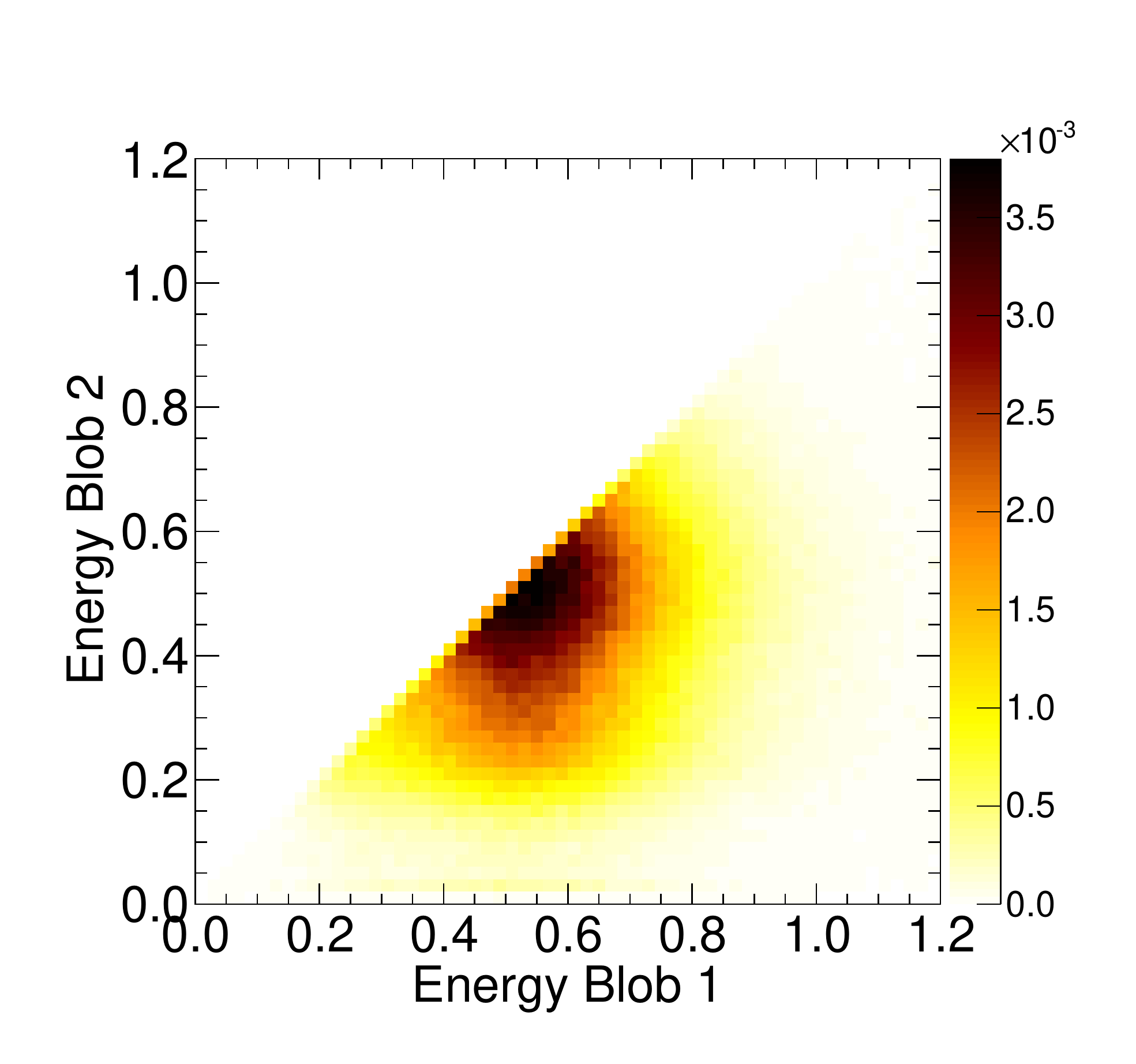}
\includegraphics[width=0.45\textwidth]{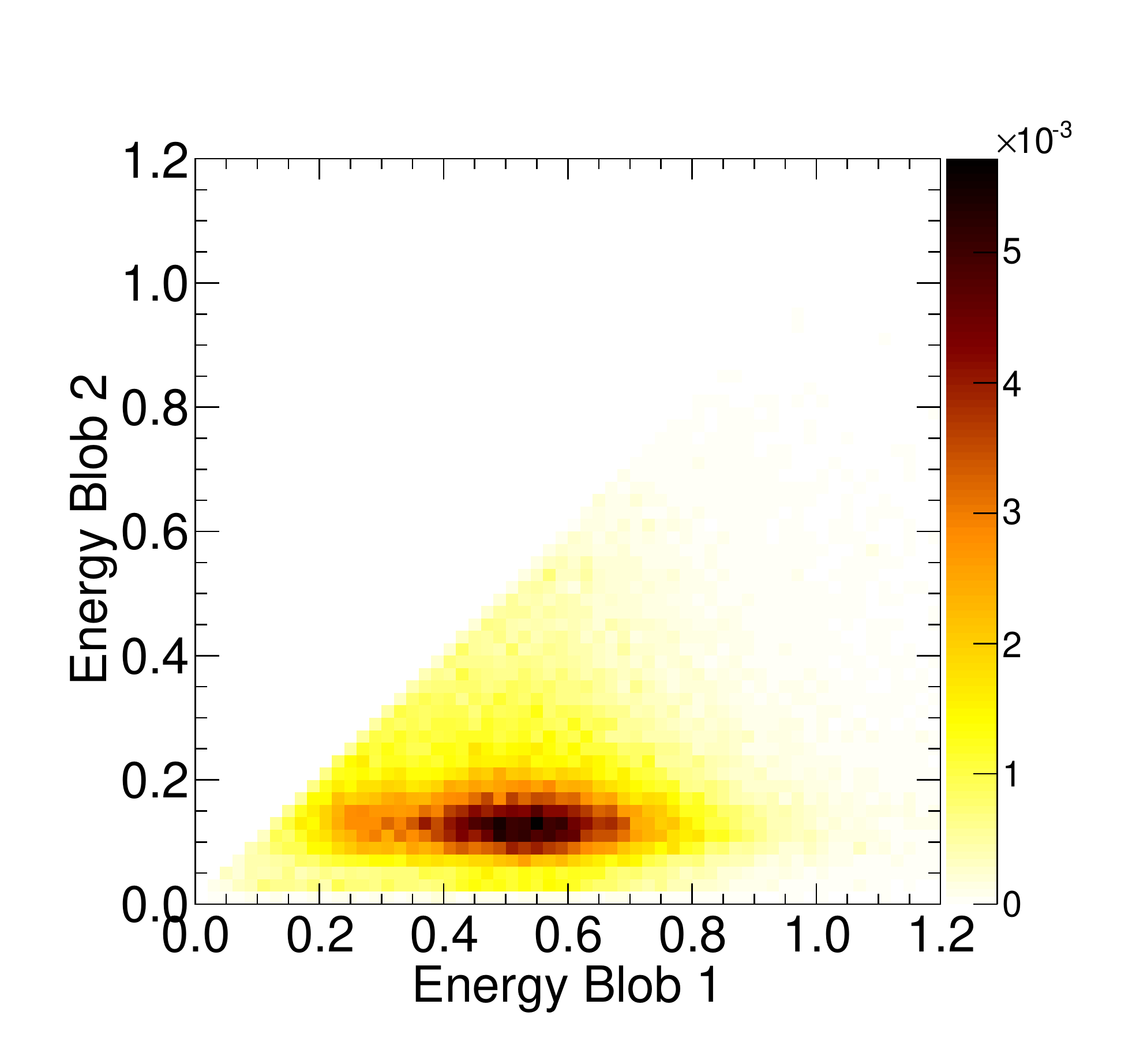}
\caption{Top: Monte Carlo simulation of signal (left) and background (right) events in xenon gas at 15~bar. The color codes energy deposition in the gas. The signal tracks consist of two electrons emitted from a common vertex, and thus they feature blobs at both ends. Background event are, typically, single-electron tracks, featuring only one blob. Bottom: Probability distribution of signal (left) and background (right) events in terms of the energies of the end-of-track blobs. The blob labelled as `1' corresponds to the more energetic one, whereas `blob 2' corresponds to the less energetic of the two.} \label{fig.ETRK2}
\end{figure}
%%%%%

In addition of excellent energy reconstruction, NEXT has a topological signature, not available in most \bbonu\ detectors. 
Double beta decay events leave a distinctive topological signature in HPXe: a continuous track with larger energy depositions (\emph{blobs}) at both ends due to the Bragg-like peaks in the d$E$/d$x$ of the stopping electrons (figure \ref{fig.ETRK2}, top left). In contrast, background electrons are produced by Compton or photoelectric interactions, and are characterised by a single blob and, often, by a satellite cluster corresponding to the emission of 30-keV fluorescence X-rays by xenon (figure \ref{fig.ETRK2}, bottom left).
Reconstruction of this topology using the tracking plane provides a powerful means of background rejection, as can be observed in the figure. In our TDR we chose a conservative cut to separate double--blob from single--blob events which provided a suppression factor of 20 for the background while keeping 80\% of the signal.  DEMO has reconstructed single electrons from \NA\ and \CS\ sources, as well as double electrons from the double escape peak of \TL\, demonstrating the robustness of the topological signal.

%%%%%%%%%%%%%%%%%%%%%%%%%%%%%%%%%%%%%%%%
\subsection{SNO+} \label{subsec:SNO+}
%%%
SNO+, the follow-up of the \emph{Sudbury Neutrino Observatory} (SNO) \cite{Boger:1999bb}, is a multipurpose liquid scintillator experiment housed in SNOLAB (Ontario, Canada). The detector reuses many of the components of its predecessor, replacing the heavy water by 780~tonnes of liquid scintillator in order to obtain a lower energy threshold. The detector consists of a 12~m diameter acrylic vessel surrounded by about 9500 8-in photomultiplier tubes that provide a 54\% effective photocathode coverage. The acrylic vessel is immersed in a bath of ultra pure water that fills the remaining extent of the underground cavern, attenuating the background from external media such as the PMTs and surrounding rock. The density of the liquid scintillator (0.86~g/cm$^{3}$) being lower than that of the surrounding water leads to a large buoyant force on the acrylic vessel. To keep it in place, a hold-down rope net has been installed over the detector and anchored to the cavity floor.

The physics program of SNO+ includes the search for neutrinoless double beta decay in \TE, which will be loaded into the liquid scintillator in the form of (non-enriched) telluric acid. A loading of 0.3\%, equivalent to 780~kg of \TE, is planned for the first phase of the experiment, which will start towards the end of 2015 or beginning of 2016. 

The energy resolution of the SNO+ detector is expected to be 10.5\% FWHM at the $Q$ value of \TE\ \cite{Biller:2014eha}. Consequently, the \bbtnu\ spectrum will be an important source of background. The expected levels of uranium and thorium in the liquid scintillator can also result in substantial activity near the \bbonu\ endpoint, mostly from the decays of $^{214}$Bi and $^{212}$Bi. Nevertheless, these can be, in principle, actively suppressed via Bi-Po $\alpha$ tagging \cite{Biller:2014eha}. External backgrounds (not originating in the liquid scintillator) can be suppressed with a tight fiducial volume selection, which will cut, however, about 70--80\% of the signal. 

%%%%%%%%%%%%%%%%%%%%%%%%%%%%%%%%%%%%%%%%
\subsection{SuperNEMO} \label{subsec:SuperNEMO}
\begin{figure}
\centering
\includegraphics[trim=220 155 220 155, clip, width=0.75\textwidth]{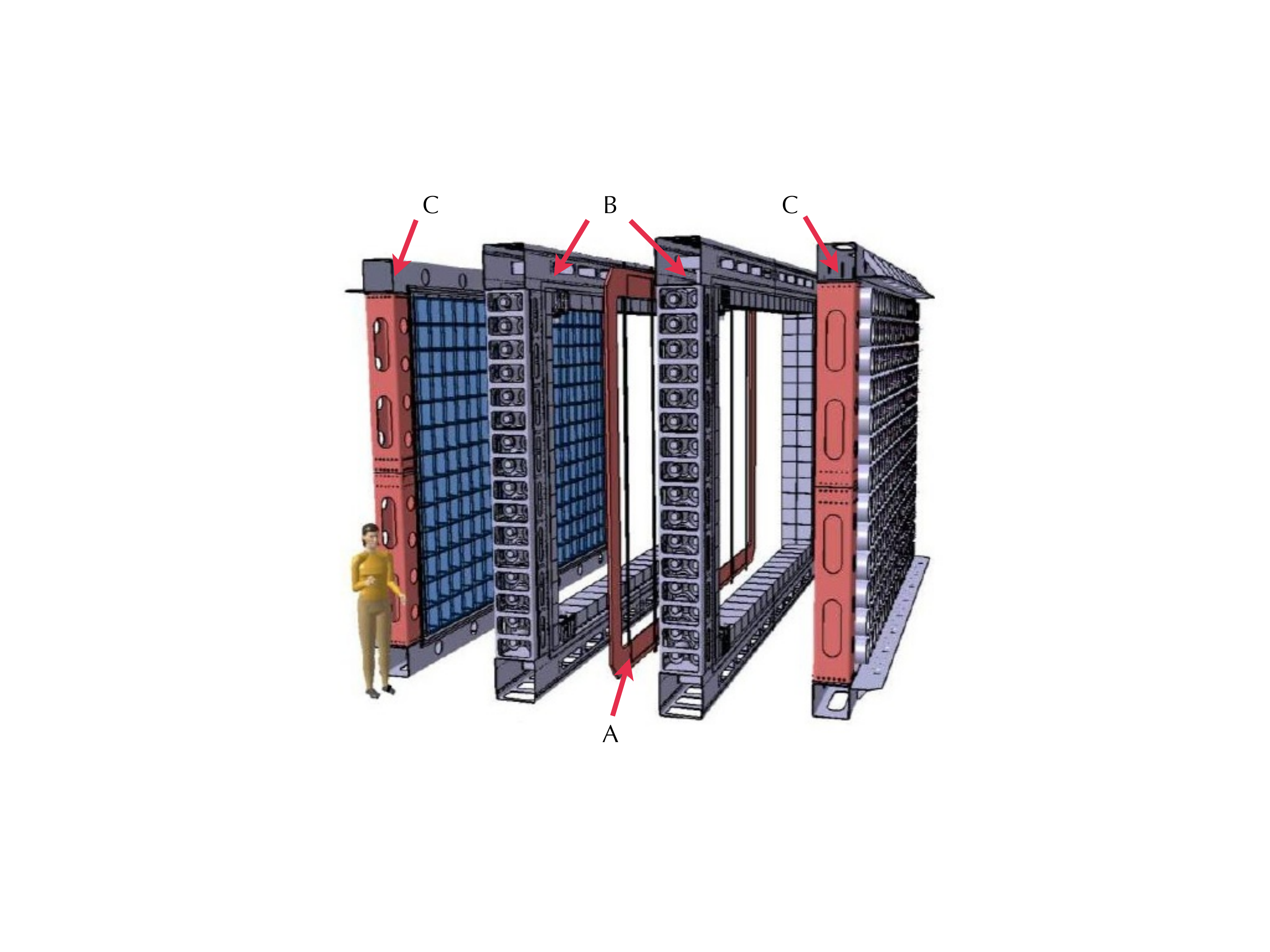}
\caption{Exploded view of a SuperNEMO module. The source foil (A), with approximately 5~kg of \SE, is sandwiched between two tracking chambers (B) and two calorimeter walls (C).} \label{fig:SuperNEMO}
\end{figure}

%%%
The SuperNEMO collaboration proposes the construction of up to 20 modules 6.2~m long, 4.1~m high and 2.1~m wide, each one containing a thin source foil of 5--7~kg of \SE\ (\ND\ and \CA\ are also contemplated, in case their isotopic enrichment becomes feasible) \cite{Lang:2013fta}. A drawing of a SuperNEMO module can be seen in Figure~\ref{fig:SuperNEMO}. Each module will consist of a central source foil, 3~m long and 4.5~m high, with a surface density of about 40~mg/cm$^{2}$, placed in between two tracking chambers --- drift cells operating in Geiger mode --- surrounded by calorimeter walls made of plastic scintillator blocks coupled to PMTs.

%%%%%%%%%%
%%%%%%%%%%

The SuperNEMO modules expect to improve the performance of the NEMO detector, both at the level of energy resolution (the projected resolution is 4\% FWHM at 3~MeV, almost a factor of 2 better than in NEMO-3) and background rate. Indeed, the 
projected background rate in the energy region around \Qbb\ is $5\times10^{-5}$~\ckky, about 25 times better than in NEMO-3, requiring \BI\ and \TL\ impurities in the source foils to be reduced to less than $2~\mu$Bq/kg and $10~\mu$Bq/kg, respectively. Radon, which will deposit \BI\ in the foil must also be controlled accordingly. 

The SuperNEMO collaboration is currently building a module, 
called the \emph{Demonstrator}, which will house a slightly denser source of approximately 7~kg of \SE. The beginning of operations is projected for 2015. It will be installed in the former location of the NEMO-3 apparatus.

\section{Towards the tonne scale} \label{tonne}

%%%%%%%%%%%%%%%%%%%%%%%%%%%%%%%%%%%%%%%%%%%%%%%%%%%%%%%%%%%%
\subsection{Reach of the current generation of experiments}
%%%
The importance of the worldwide experimental program searching for neutrinoless double beta decay can hardly be overstated: the discovery of the radioactive process is, perhaps, the only way of determining the nature of neutrino mass and proving the violation of total lepton number. The current generation of experiments consists of projects using source masses of tens to hundreds of kilograms and a variety of detection techniques.  Three of these experiments (EXO-200, KamLAND-Zen and GERDA-I) are operating already and have released physics results, excluding the lifetimes below $10^{25}$~years, with typical exposures of 100 kg$\cdot$ year for the xenon-based experiments, and about  20 kg$\cdot$ year for GERDA-I. In the next 2-3 years, GERDA-II will add 20 kg of new detectors, while {\scshape Majorana}, based also in germanium diodes will start operation with an additional mass of 30 kg. Thus, an exposure in the range of some 200 kg$\cdot$ year, will be within the reach of germanium-based detectors.  NEXT-100  will start operation in 2017 and its initial run foresees also reaching exposures 
of $\sim$ 200 kg $\cdot$ year. Similar, or even larger exposures will be obtained by EXO-200, KamLAND-ZEN, and SNO+. The SuperNEMO demonstrator will deploy a mass of 7 kg, and thus will reach a exposure in the range of $\sim$  20 kg$\cdot$ year.. 

%%%%%%%%%%
\begin{table}[!]
\centering
\caption{Basic operational parameters of the \bbonu-decay experiments of the current generation: \bb\ source isotope, energy resolution (FWHM), $\Delta E$; background rate in the region of interest around \Qbb; signal detection efficiency, $\varepsilon$; and source mass.} \label{tab:ParamsCurrentGeneration}
\small
\begin{tabular*}{\textwidth}{@{\extracolsep{\fill}} l c D{.}{.}{3.0} D{.}{.}{0.5} c D{.}{.}{3.0}}
\toprule
Experiment & Isotope & \multicolumn{1}{c}{$\Delta E$} & \multicolumn{1}{c}{Bkgnd.\ rate} & \multicolumn{1}{c}{$\varepsilon$} & \multicolumn{1}{c}{Mass} \\
           &         & \multicolumn{1}{c}{(keV)} & \multicolumn{1}{c}{(keV$^{-1}$~kg$^{-1}$~yr$^{-1}$)} & \multicolumn{1}{c}{$(\%)$} & \multicolumn{1}{c}{(kg)} \\ \midrule
CUORE-0\hair\textsuperscript{\itshape a} \cite{Giachero:2014hva} & \TE\ & 5 & 0.23 & 78 & 11 \\
CUORE\hair\textsuperscript{\itshape b} \cite{Alessandria:2011rc} & \TE\ & 5 & 0.04 & 87 & 206 \\
GERDA-I\hair\textsuperscript{\itshape a} \cite{Agostini:2013mzu} & \GE\ &   5 & 0.013 & 62 & 15 \\
GERDA-II\hair\textsuperscript{\itshape b} \cite{Macolino:2014vya} & \GE\ & 3 & 0.001 & 66 & 33 \\
EXO-200\hair\textsuperscript{\itshape a} \cite{Albert:2014awa} & \XE\ & 88 & 0.002 & 85 & 76 \\
KamLAND-Zen\hair\textsuperscript{\itshape a} \cite{Gando:2012zm, Asakura:2014lma} & \XE\ & 243 & 0.00014  & 25 & 348 \\
{\scshape Majorana}\textsuperscript{\itshape c} \cite{Abgrall:2013rze} & \GE\ & 4 & 0.0009 & 70 & 25 \\
NEXT-100\hair\textsuperscript{\itshape c} \cite{Gomez-Cadenas:2013lta} & \XE\ &  18 & 0.0006  & 28 & 91 \\
SNO+\hair\textsuperscript{\itshape c} \cite{Biller:2014eha} & \TE\ & 264 & 0.0001  & 15 & 800 \\
SuperNEMO-D\hair\textsuperscript{\itshape c} \cite{Guzowski:2014ina} & \SE\ & 120 & 0.0005 & 30 & 7 \\ 
\bottomrule \\[-8pt]
\multicolumn{6}{l}{\textsuperscript{\itshape a} \footnotesize The experiment is running and has measured its operational parameters.} \\[-2pt]
\multicolumn{6}{l}{\textsuperscript{\itshape b} \footnotesize The experiment has proven its feasibility with a demonstrator.} \\[-2pt]
\multicolumn{6}{l}{\textsuperscript{\itshape c} \footnotesize The operational parameters are estimations based on R\&D results and simulations. } \\
\end{tabular*}
\end{table}
%%%%%%%%%%

The basic operational parameters of all the experiments of the current generation are listed in Table~\ref{tab:ParamsCurrentGeneration}. The size of the uncertainties associated to the parameters varies according to the state of development of each project.  Naturally, the smallest uncertainties correspond to the running experiments, which have \emph{measured} their operational parameters. CUORE and GERDA-II have assessed their expected performance with setups operating under conditions similar to those of the final experiment. The remaining experiments base their expectations on results obtained with R\&D prototypes, ancillary measurements and Monte Carlo simulations. 

%%%%%%%%%%%
%\begin{figure}
%\centering
%\includegraphics[width=0.49\textwidth]{img/chapter-08/SensHLCurrentGenGe76.pdf}
%\includegraphics[width=0.49\textwidth]{img/chapter-08/SensHLCurrentGenSe82.pdf}
%\includegraphics[width=0.49\textwidth]{img/chapter-08/SensHLCurrentGenXe136.pdf}
%\includegraphics[width=0.49\textwidth]{img/chapter-08/SensHLCurrentGenTe130.pdf}
%\caption{Half-life sensitivity (at 90\% CL) of the current-generation experiments in terms of the exposure. The experiments are grouped according to their \bb\ source isotope. From left to right and top to bottom: \GE\ (GERDA and {\scshape Majorana}), \SE\ (SuperNEMO), \XE\ (EXO-200, KamLAND-Zen and NEXT-100) and \TE\ (CUORE and SNO+).} \label{fig:SensHLCurrentGen}
%\end{figure}
%%%%%%%%%%%

%%%%%%%%%%
\begin{figure}
\centering
\includegraphics[width=0.6\textwidth]{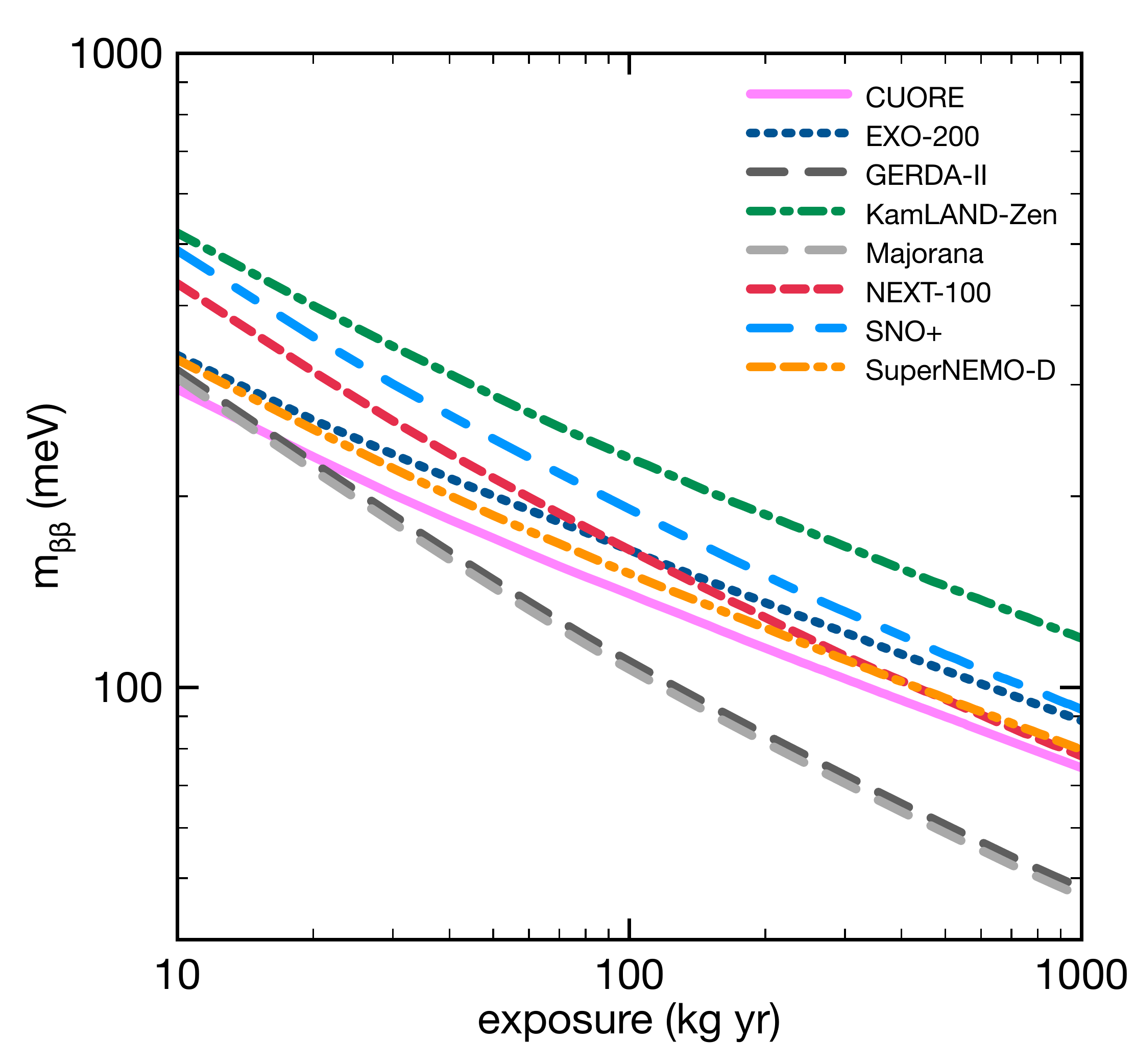}
\includegraphics[width=0.6\textwidth]{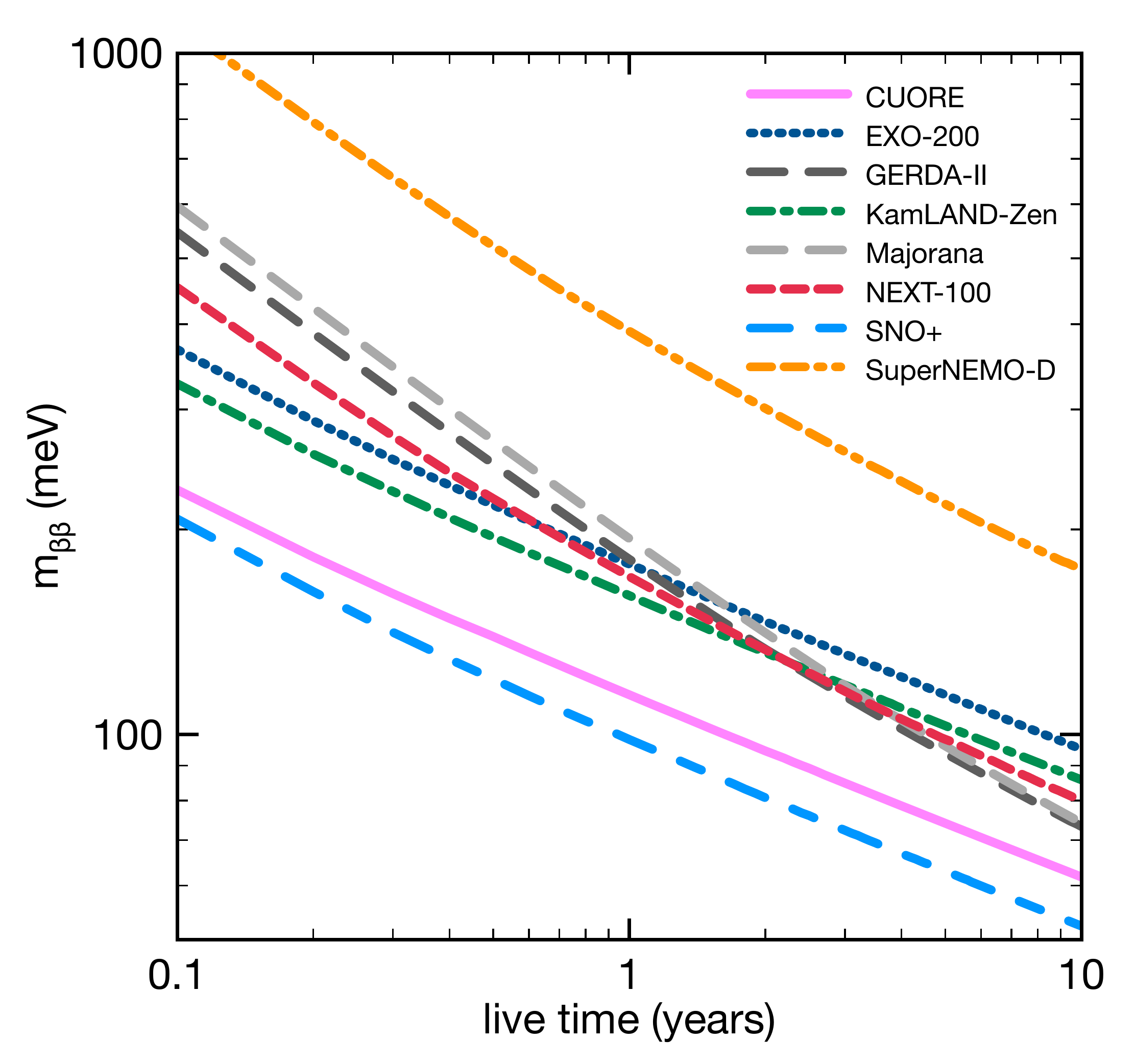}
\caption{Sensitivity to \mbb\ (at 90\% CL) of the current generation of \bbonu-decay experiments. Top: \mbb\ sensitivity as a function of the exposure. Bottom: \mbb\ sensitivity as a function of the experiment's live time.} \label{fig:SensNMCurrentGen}
\end{figure}
%%%%%%%%%%

%Figure~\ref{fig:SensHLCurrentGen} shows the half-life sensitivity (at 90\% CL) of the current generation of \bbonu-decay experiments, calculated using the parameters of Table~\ref{tab:ParamsCurrentGeneration} and the method described in \textsection~\ref{sec:SensitivityDef}. The experiments are grouped according to their chosen source isotope. 
%
%In Figure~
Figure~\ref{fig:SensNMCurrentGen} (top panel) shows the sensitivity to the effective neutrino Majorana mass, \mbb (at 90\% CL) of the current generation of \bbonu-decay experiments, calculated using the parameters of Table~\ref{tab:ParamsCurrentGeneration} and the IBM-2 nuclear matrix elements \cite{Barea:2013bz}. For the same exposure, the \GE-based experiments ---\hair GERDA-II and {\scshape Majorana}\hair--- would be the most sensitive detectors of the current generation if they attained their target operational parameters. In its first phase, GERDA has achieved a background rate of about $10^{-2}$~keV$^{-1}$~kg$^{-1}$ \cite{Agostini:2013mzu}. A reduction of a factor of 10 is expected for the second phase of the experiment thanks to the use of broad-end germanium detectors ---\hair which present improved pulse shape discrimination and energy resolution\hair--- and the instrumentation of the liquid argon bath to veto external backgrounds \cite{Macolino:2014vya}. The {\scshape Majorana} Collaboration pursues a slightly better background rate than that of GERDA-II with a more conventional design for the detector, paying particular attention to the selection of radiopure materials. For instance, the required purity levels for the electroformed copper used in the inner layers of shielding and in the detector holders are extremely stringent, at the level of $0.3~\mu$Bq/kg of $^{232}$Th or $^{238}$U or below \cite{Abgrall:2013rze}. %The case of SuperNEMO is significantly more uncertain than the previous two. The Collaboration expects to improve the background rate achieved in NEMO-3, the predecessor of SuperNEMO, by approximately a factor of 25 by reducing the contaminants in the source foil below a few $\mu$Bq/kg and the activity of radon in the tracking chamber from 5~mBq/m$^{3}$ to 0.15~mBq/m$^{3}$ \cite{Guzowski:2014ina}.

\begin{figure}
\centering
\includegraphics[width=0.7\textwidth]{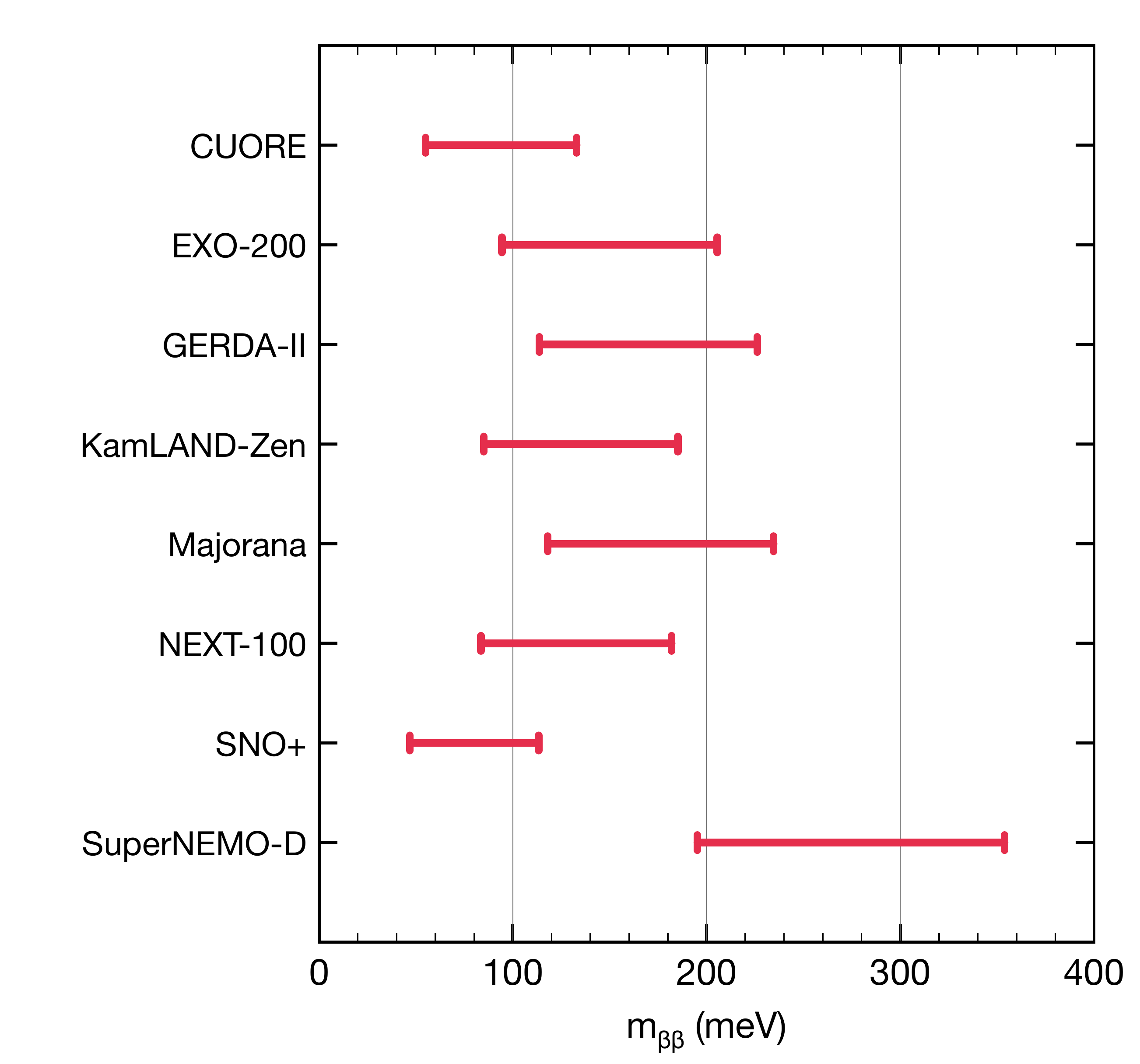}
\caption{Effective neutrino Majorana mass (\mbb) sensitivity ranges (at 90\% CL) defined by the largest and smallest NME calculations for a 3-year run of the \bbonu-decay experiments of the current generation.} \label{fig:SensNMCurrentGen3YearsRun}
\end{figure}

The experiments of the current generation deploy very different source masses, and hence they cannot reach the same exposures with equal ease. To take this into account, we have represented in the bottom panel of Fig.~\ref{fig:SensNMCurrentGen} the \mbb\ sensitivity as a function of the experiments' live time. In this case, the experiments reaching the best sensitivities are those with the highest effective source masses (i.e.\ mass times detection efficiency): SNO+ and CUORE. These \TE-based experiments would achieve a sensitivity to \mbb\ below 100~meV after a 3-year run, with the \GE\ and \XE-based experiments reaching \mbb\ sensitivities about a 30\% higher. Nevertheless, the uncertainties deriving from the NMEs blur significantly this picture, as shown in Figure~\ref{fig:SensNMCurrentGen3YearsRun}, where we represent the \mbb\ sensitivity ranges defined for each experiment by the largest and smallest NME calculations after a run of 3~years.

%%%%%%%%%%
%%%%%%%%%%

The reliability of the sensitivity estimates given above depends, of course, on how realistic the operational parameters of each experiment are. Besides, in our sensitivity computation, we have neglected systematic uncertainties and any energy-shape information that may be present in the energy distribution of events. Systematic uncertainties may possibly affect the parameters listed above, especially the knowledge of the backgrounds, and deteriorate the sensitivity. On the other hand, use of additional information beyond the overall count rate of \bbonu-decay candidates within the ROI may yield some sensitivity improvement. While important, both effects would be extremely difficult to incorporate in such a sensitivity comparison, given that most current-generation experiments discussed here have not even started their commissioning phase yet.

In summary, most of the experiments appear to have a a very good chance to reach a sensitivity of 100 meV or better after a few years of effective live time. Given the uncertainties, we cannot predict which among the different experiments will provide the best sensitivity after, say, a 3-year run. To this end, a better knowledge of the actual values for the background rates, of the systematic uncertainties affecting the measurement, and of the NMEs would be necessary for all experiments. The \GE\ claim should be unambiguously solved by current-generation experiments using different isotopes, but it appears that it will be almost impossible for this generation to discover \bbonu-decay if the neutrino mass spectrum is hierarchical rather than degenerate, as favoured by the cosmological constrains on the lightest neutrino mass.

%%%%%%%%%%%%%%%%%%%%%%%%%%%%%%%%%%%%%%%%%%%%%%%%%%%%%%%%%%%%
\subsection{Exploring the inverted-hierarchy region of neutrino masses}
%%%
We turn now our attention to the future prospects of the field. The goal of the next generation of \bbonu-decay experiments is to probe effective Majorana neutrino masses down to 10--20~meV, fully covering the inverted hierarchy region. Such future searches will involve detectors at the tonne or multi-tonne scale in \bb\ isotope mass. Most collaborations are discussing already possible designs for a tonne-scale version of their experiments. Nevertheless, the diversity of experimental approaches we are currently witnessing will not be viable in the next generation given the cost and difficulty associated; only a few approaches ---\hair most likely based on different isotopes\hair--- are going to be retained. 

%%%%%%%%%%
\begin{figure}
\centering
\includegraphics[width=0.45\textwidth]{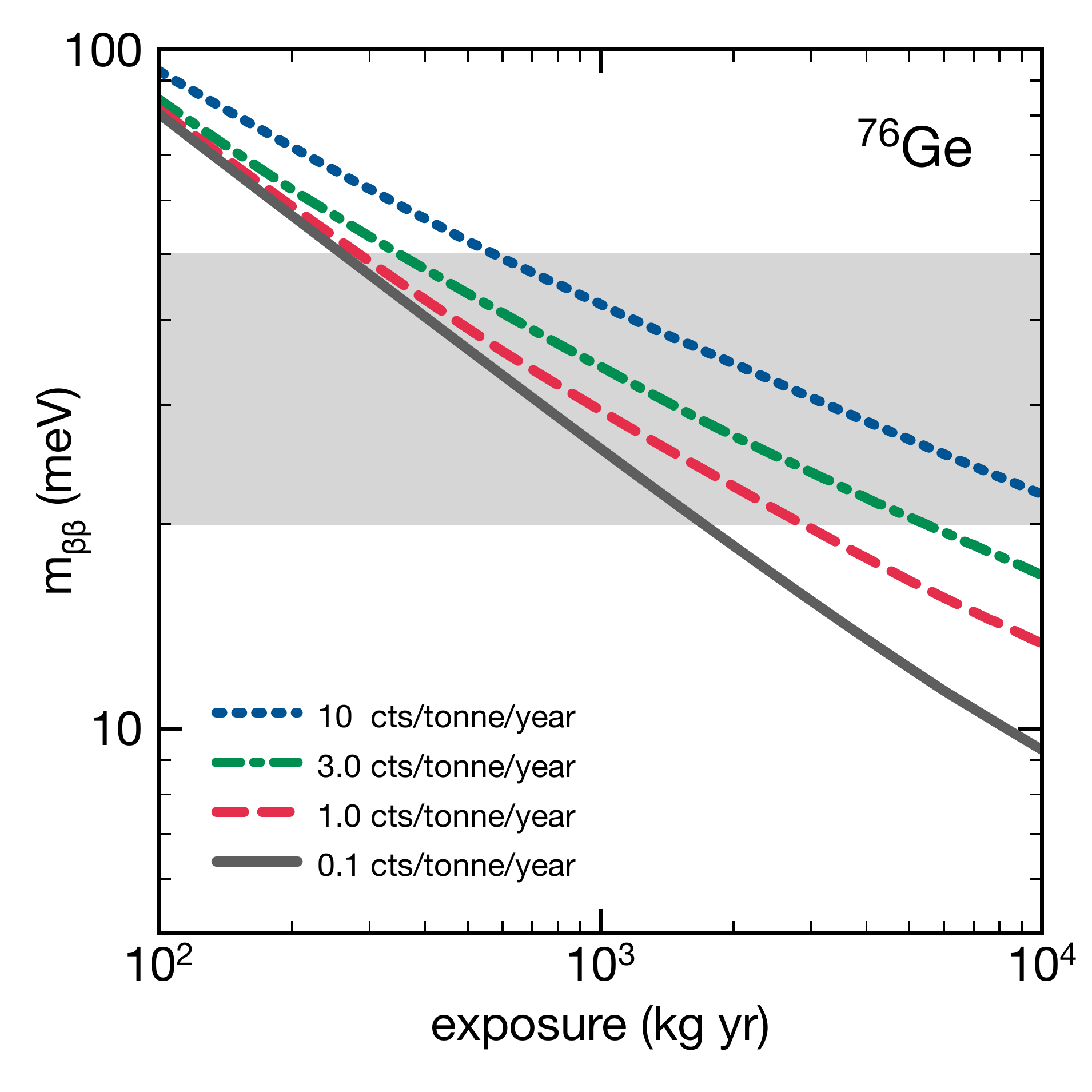}
\includegraphics[width=0.45\textwidth]{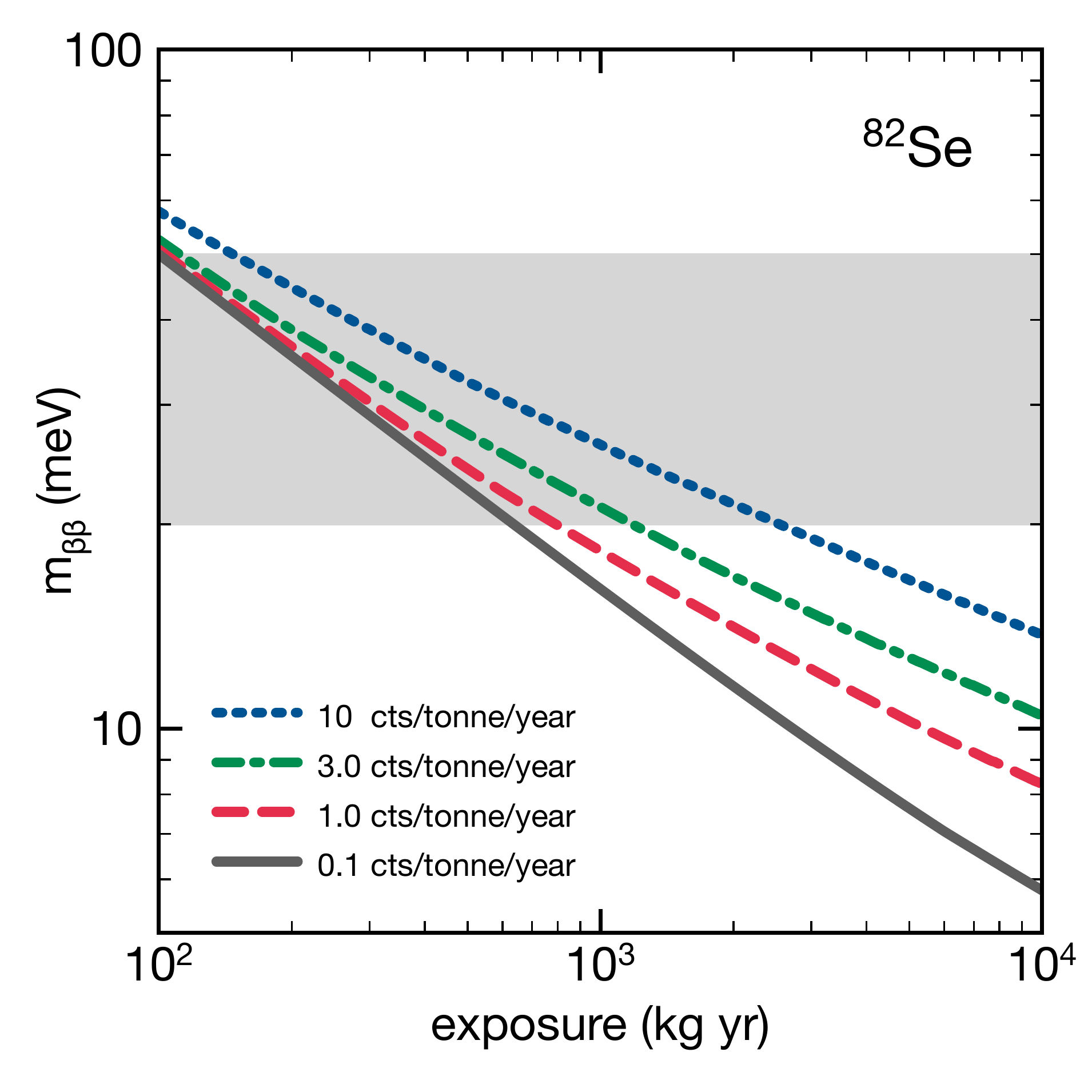}
\includegraphics[width=0.45\textwidth]{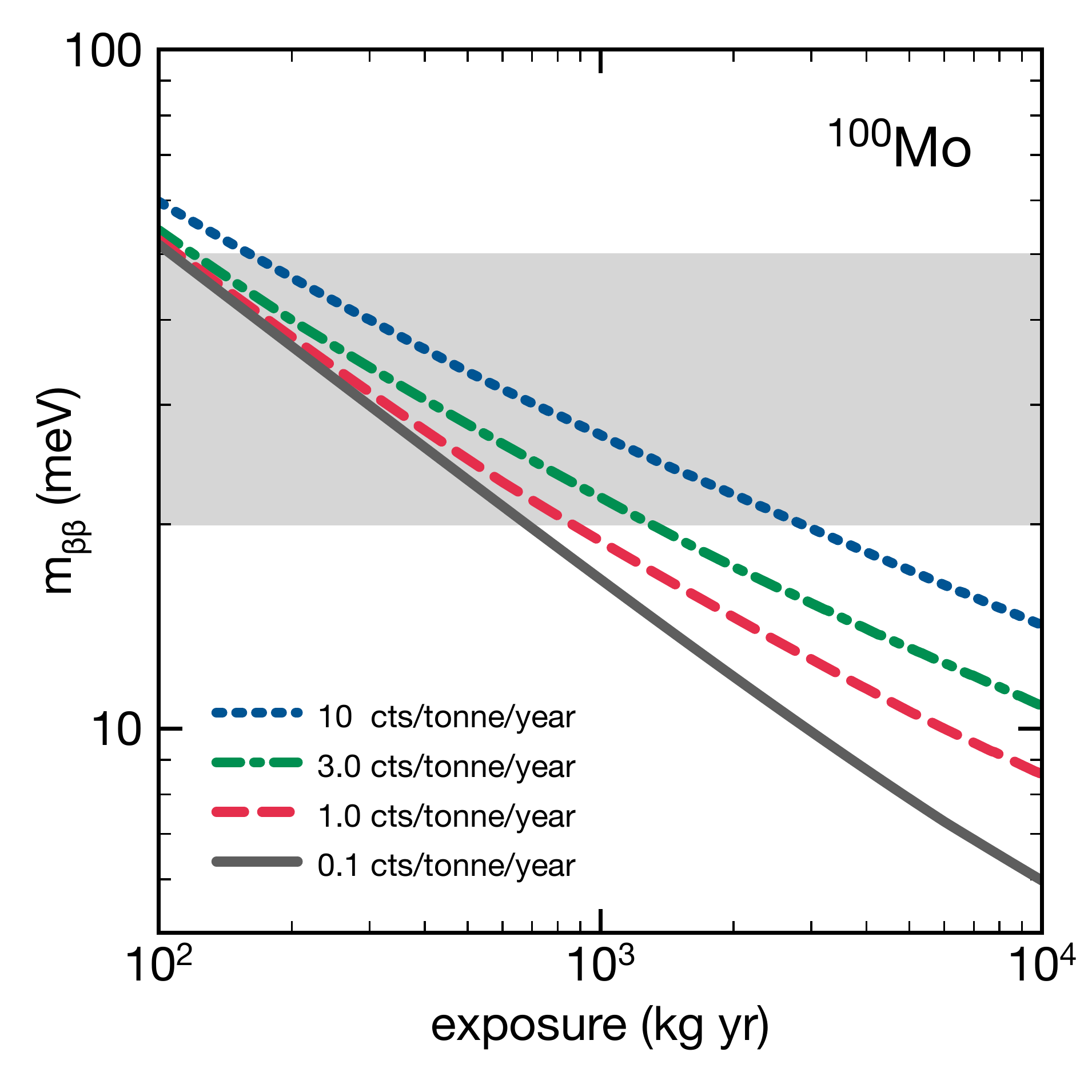}
\includegraphics[width=0.45\textwidth]{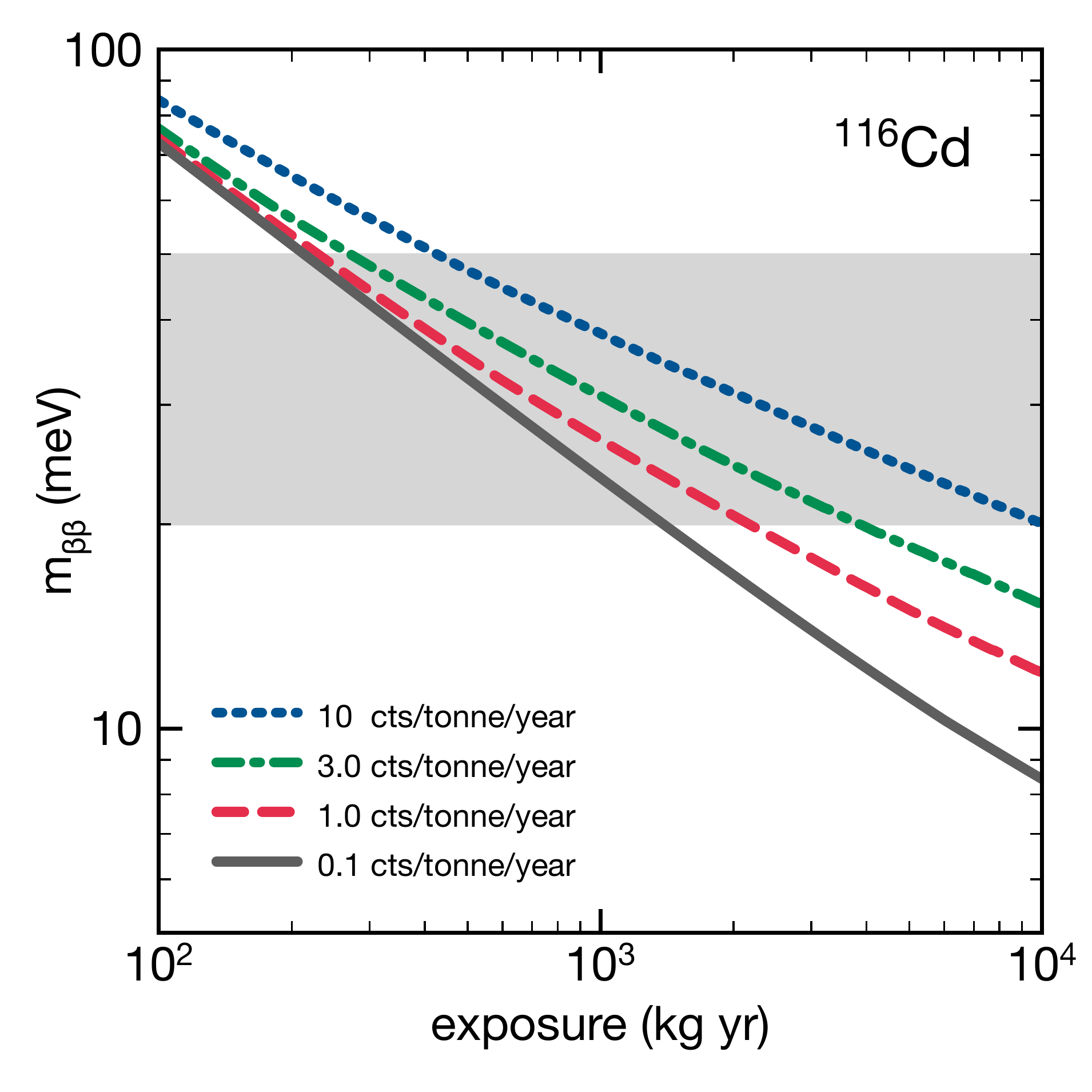}
\includegraphics[width=0.45\textwidth]{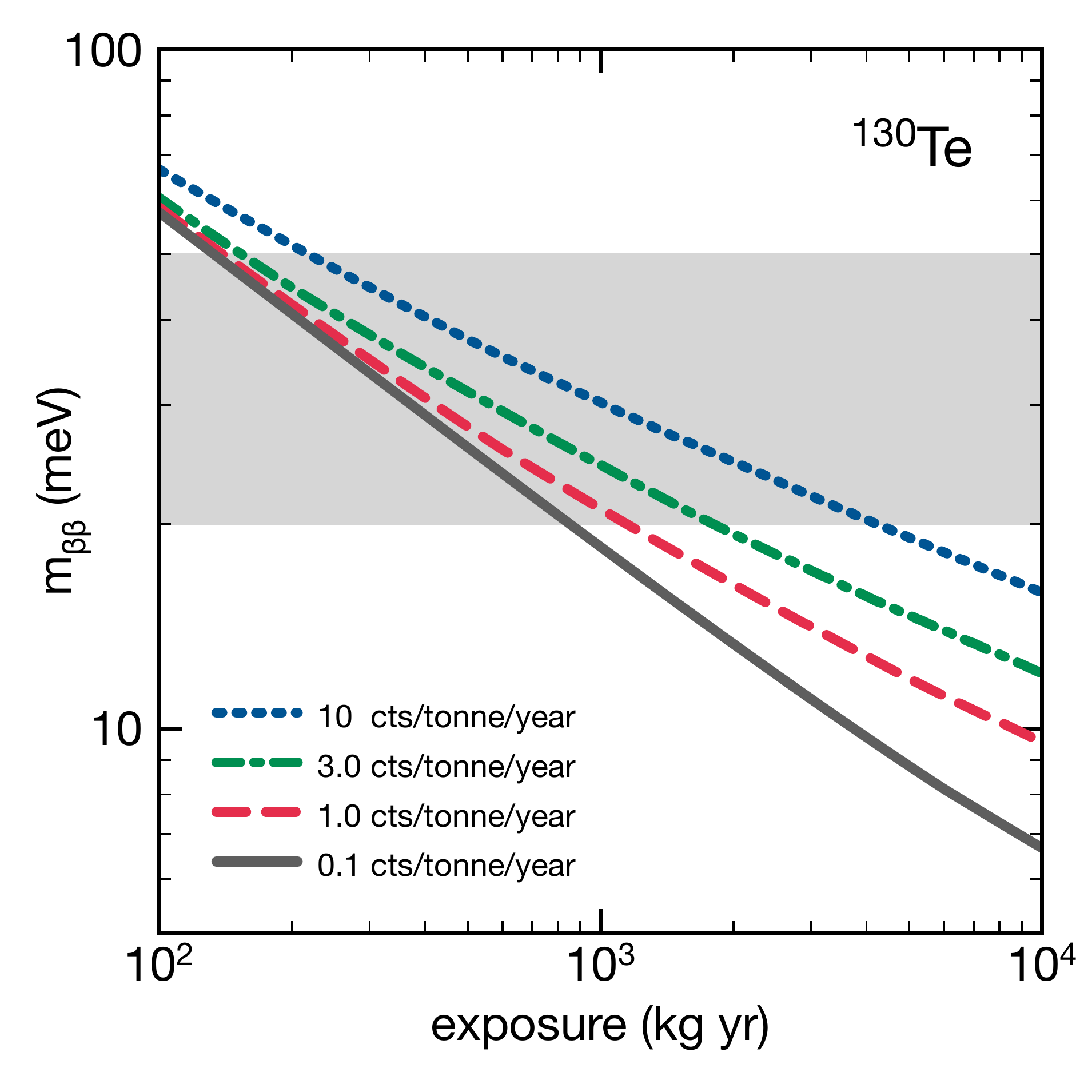}
\includegraphics[width=0.45\textwidth]{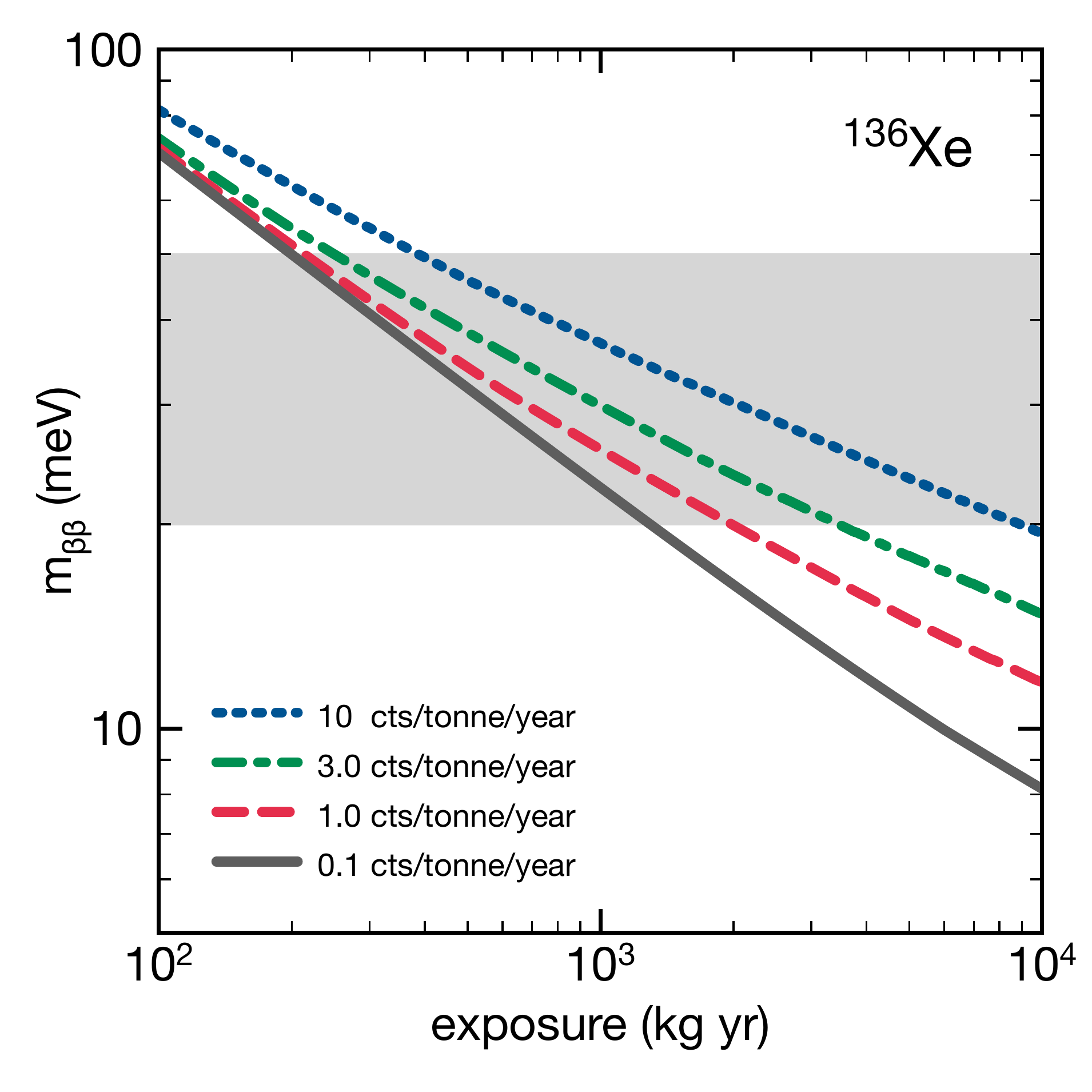}
\caption{Exposure dependence of the \mbb\ sensitivity (at 90\% CL) of perfectly-efficient experiments based on 6 different isotopes and with 4 different assumptions for the background rate in the ROI. The grey band represents the inverted-hierarchy region of neutrino masses.} \label{fig:FutureGen}
\end{figure}
%%%%%%%%%%

Figure~\ref{fig:FutureGen} shows the \mbb\ sensitivity of perfectly-efficient experiments based on six different \bb\ isotopes (\GE, \SE, \MO, \CD, \TE\ and \XE) and for four different assumptions for the background rate in the ROI. These graphs let us estimate the requirements that next-generation experiments would have to fulfil in order to meet their physics goal. We discuss them briefly in the following:
%%%
\begin{itemize}
\item GERDA and {\scshape Majorana} plan to merge their efforts towards a tonne-scale germanium experiment. In order to reach in a reasonable time \mbb\ sensitivities in the range 10--20~meV, their successor experiment would require at least a tonne of enriched material, i.e.\ 20--25 times more mass than that deployed at present by the two experiments. Besides, the background rate would have to be reduced by a factor of 10 with respect to the target rate of the current-generation experiments, reaching $10^{-4}$~counts/(keV~kg~year).
\item A tonne-scale version of CUORE would require isotopically-enriched TeO$_{2}$ bolometers and a background rate two orders of magnitude smaller than the one expected in its present version. The detection of the Cherenkov light produced by signal events in the crystals could be used to suppress the main source of background in CUORE ($\alpha$ activity on the surface of the bolometers) \cite{TabarellideFatis:2009zz, Casali:2014vvt}. Alternatively, the LUCIFER \cite{Beeman:2013sba}, LUMINEU \cite{Barabash:2014una} and AMoRE \cite{Bhang:2012gn} projects are exploring the use of scintillating bolometers (Zn$^{100}$MoO$_{4}$, $^{40}$Ca$^{100}$MoO$_{4}$, $^{116}$CdWO$_{4}$ or Zn$^{82}$Se), which offer an additional experimental signature for background suppression, but a slightly worse energy resolution. The procurement and enrichment of these crystals would also be notably more expensive than for the \TE-based ones. The ultimate sensitivity of bolometers may be limited by the accidental pile-up between multiple \bbtnu\ events in the same crystal due to their slow response \cite{Artusa:2014wnl}. 
\item The two large liquid-scintillator calorimeters, SNO+ and KamLAND-Zen, are planning future phases for the exploration of the inverse neutrino hierarchy. KamLAND-Zen intends to dissolve about 1000~kg of enriched xenon in the liquid scintillator. SNO+ can, in principle, increase the concentration of telluric acid in the liquid scintillator by a factor of 10 without affecting the detection and optical properties of the mixture \cite{Biller:2014eha}. The energy resolution of both detectors may be improved from the current 10\% FWHM to values close to 6\% FWHM with the use of a brighter liquid scintillator and an enhancement of the light collection efficiency. As regards the background level, SNO+ would have to suppress by some means its present target rate, $10^{-4}$~counts/(keV~kg~year), by an order of magnitude. KamLAND-Zen would require an even higher suppression factor to compensate for the lower deployed mass and the differences between the \bb\ sources. The sensitivity of these experiments will be limited, ultimately, by irreducible backgrounds such as the \bbtnu\ spectrum or the solar neutrino flux.
\item The EXO Collaboration has started the design of a 5-tonne liquid xenon TPC called nEXO. In order to fully probe the inverted-hierarchy region, nEXO would require a background rate approximately two orders of magnitude better than that achieved in EXO-200. Self-shielding would help suppressing the external backgrounds, even though at the cost of reducing the fiducial mass. In addition, the detector will require neutron shielding to mitigate the activation of \XE.
\item The NEXT Collaboration is considering a xenon gas detector with a mass in the range of 1 to 3 tonnes, energy resolution close to 0.5\% FWHM at 2.5~MeV, and a background rate of the order of $5\times10^{-5}$ counts/(keV~kg~yr). Besides possible improvements in the radioactivity budget of the detector (more radiopure photosensors, for example), the target background rate can be attained enhancing the discrimination power of the tracking signature with a moderate magnetic field. A single energetic electron should produce a clear single spiral with radius indicative of its momentum, and a double-electron track with the same energy will produce two spirals each with much less momentum and originating from a common vertex. This information provides an additional way of separating single-electrons arising from background processes from double electrons produced in \bbonu\ decays, in spite of the large multiple scattering that the electrons suffer in the dense gaseous xenon.
\item SuperNEMO has, by far, the worst mass-to-volume ratio among the techniques considered for \bbonu-searches: a SuperNEMO module houses 5--10 kg of \bb\ isotope in approximately 53~m$^{3}$, whereas, for example, the NEXT-100 pressure vessel, with a volume of about 3~m$^{3}$, contains 100~kg of enriched xenon in its active volume. Considering this and the limited underground space available, a tonne-scale version of the SuperNEMO experiment seems very unlikely, if not impossible.
\end{itemize}
%%%

\section{Conclusions} \label{Conclusions}
%%%
At the turn of the 21st century, the observation of neutrino oscillations, which implies that neutrinos are massive particles, and the possible evidence of \bbonu-decay in the Heidelberg-Moscow experiment \cite{KlapdorKleingrothaus:2001ke} boosted the interest in neutrinoless double beta decay searches, prompting a new generation of experiments characterized by source masses in the range of tens to hundreds of kilograms and based on a variety of detection techniques. Indeed, searching for neutrinoless double beta decay is well motivated: first, there is no fundamental reason why total lepton number should be conserved; and second, Majorana neutrinos provide natural explanations for both the smallness of neutrino masses and the baryon number asymmetry observed in the Universe. Consequently, theoretical prejudice in favour of a Majorana nature for neutrinos has gained widespread consensus.

In this paper, we have review the main experiments of the new-generation experiments. We have made an attempt at a quantitative comparison of the physics case of the different \bbonu-decay experiments of the current generation. There is an intense competition among these experiments to establish themselves as the best approach for neutrinoless double beta decay searches. Since the exploration of the inverted-hierarchy region will unavoidably need to involve experiments at the tonne or multi-tonne scale in isotope mass, the current experiments do not only have to proof their performance for source masses of the order of 100~kg, but also show their scalability to the tonne scale, including the ability to improve their current background rate by, at least, a factor of 10. It is, probably too early to decide on the most promising technologies, in particular if one adds the constrain of cost, and feasibility to the physics requirements. During the next five years or so, CUORE, SNO+, NEXT-100 and the SuperNEMO demonstrator will add their contribution to the exploration of the Majorana landscape already initiated by GERDA, EXO-200 and KamLAND-ZEN. If the effective neutrino mass lies in the degenerated regime, up to some 100 meV, a discovery can be made. Else, the field will have, presumably, matured enough, to choose the most promising paths toward the full exploration of the Majorana landscape.

\acknowledgments{We gratefully acknowledge the support of the \emph{European Research Council} (ERC) under the Advanced Grant 339787-NEXT and the \emph{Ministerio de Econom\'ia y Competitividad} of Spain under grants CONSOLIDER-Ingenio 2010 CSD2008-0037 (CUP), FPA2009-13697-C04 and FIS2012-37947-C04. We thank G.~Senjanovic and C.~Pe\~na Garay for comments and suggestions.}

\end{document}